\documentclass[aps,prx,twocolumn,showpacs]{revtex4-1}
\pdfoutput=1
\usepackage{hyperref}
\hypersetup{
colorlinks = true, linkcolor = [rgb]{0,0,1}, citecolor = [rgb]{0,0,1},
                            urlcolor = [rgb]{0,0,1}}
\usepackage{graphicx,epsf}
\usepackage{amsmath}
\usepackage{amssymb}
\usepackage{color}
\usepackage{esint}
\usepackage{wasysym}

\newcommand{\bB}{{\bf B}}

\newcommand{\bq}{{\bf q}}

\newcommand{\bp}{{\bf p}}
\newcommand{\bk}{{\bf k}}

\newcommand{\taub}{\mbox{\boldmath $\tau $}}
\newcommand{\deltab}{\mbox{\boldmath $\delta $}}

\usepackage[normalem]{ulem} 
\newcommand\redout{\bgroup\markoverwith
{\textcolor{red}{\rule[.5ex]{2pt}{0.4pt}}}\ULon}

\begin{document}

\title{Flat-band ferromagnetism and spin waves in the Haldane-Hubbard model}

\author{Leonardo S. G. Leite  and R. L. Doretto}
\affiliation{Instituto de F\'isica Gleb Wataghin,
                  Universidade Estadual de Campinas,
                  13083-859 Campinas, SP, Brazil}

\date{\today}

\begin{abstract}
We study the flat-band ferromagnetic phase of the Haldane-Hubbard
model on a honeycomb lattice within a bosonization scheme for
flat-band Chern insulators, focusing on
the calculation of the spin-wave excitation spectrum.
We consider the Haldane-Hubbard model with the noninteracting lower
bands in a nearly-flat band limit, previously determined for the
spinless model, and at $1/4$-filling of its corresponding
noninteracting limit.  
Within the bosonization scheme, the Haldane-Hubbard model is mapped
into an effective interacting boson model, whose quadratic term allows
us to determine the spin-wave spectrum at the harmonic approximation.
We show that the excitation spectrum has two branches with a Goldstone
mode and Dirac points at center and at the $K$ and $K'$ points of the first
Brillouin zone, respectively.
We also consider the effects on the spin-wave spectrum due to an energy
offset in the on-site Hubbard repulsion energies 
and due to the presence of an staggered on-site energy term, both quantities
associated with the two triangular sublattices. In
both cases, we find that an energy gap opens at the $K$ and $K'$
points. Moreover, we also find some evidences for an 
instability of the flat-band ferromagnetic phase in the presence of the
staggered on-site energy term.
We provide some additional results for the square lattice
topological Hubbard model previous studied within the bosonization
formalism and comment on the differences between the bosonization
scheme implementation for the correlated Chern insulators on both 
square and honeycomb lattices.

\end{abstract}

\maketitle

\section{Introduction}
\label{sec:intro}

The tight-binding model on a honeycomb lattice with broken time-reversal
symmetry proposed by Haldane \cite{haldane1988model} is an interesting
example of a Chern band insulator \cite{kane13,rpp-rachel18}. 
At half-filling, it can exhibit a quantized Hall conductance in the
absence of an external magnetic field. 
This so-called anomalous quantum Hall effect \cite{qi11} 
is indeed related to the fact that the electronic band
structure of Haldane's model is topologically nontrivial, i.e., the
corresponding Chern numbers of each band are finite \cite{kane13,rpp-rachel18}.  
Interestingly, the model was experimentally implemented in a system
with ultracold fermions in an optical honeycomb lattice \cite{jotzu14} 
(see also the reviews \cite{zoller16,rmp-cooper19}).

A lot of effort has also been devoted to the study of the interplay
between the topological properties of electronic band structures and
the electron-electron interaction \cite{rpp-rachel18,assaad13}.
A particular correlated Chern insulator that has been receiving some
attention in recents years is a natural extension of Haldane's
model \cite{haldane1988model}, the so-called Haldane-Hubbard model 
\cite{he2011chiral,he2011topological,maciejko2013topological,
hickey15,hickey2016haldane,zheng15,vanhala16,wu2016quantum,arun16,troyer16}.  
Here the electronic spin is explicitly taken into account,
time-reversal symmetry is broken, and correlation effects are
described via an on-site Hubbard repulsion term 
[see Eq.~\eqref{eqHH0} below].
At half-filling, the phase diagram of the Haldane-Hubbard model has
been determined via different mean-field approaches 
\cite{he2011chiral, he2011topological,hickey15,zheng15, arun16}
and numerical methods
\cite{hickey2016haldane, vanhala16, wu2016quantum, troyer16}.
It was shown that the model supports a Chern insulator phase for weak
interactions and a trivial Néel magnetic ordered phase for strong
ones. Moreover, there are evidences for a first-order transition
between these two phases \cite{troyer16}, 
but also for the presence of a distinct phase in the intermediate
coupling region \cite{he2011chiral, he2011topological,wu2016quantum}.  
Differently from the time-reversal symmetric Kane-Mele Hubbard model
\cite{rpp-rachel18},  
the breaking of time-reversal symmetry in the Haldane-Hubbard model
yields the so-called fermion sign problem, 
limiting the use of quantum Monte Carlo simulations 
\cite{wu2016quantum,troyer16}.

Away from half-filling, the study of correlation effects in Chern
band insulators have also considered the possibility of realizing  
fractional quantum Hall phases in lattice models.
Indeed, the interest in fractional Chern insulators 
\cite{sondhi13,liu13,neupert15} was motivated by the
studies \cite{neupert2011fractional,sun2011nearly,tang2011high}, 
which showed that a series of tight-binding models with only
short-range hoppings can display nearly-flat and
topologically nontrivial electronic bands once the model parameters
are properly chosen. 
Due to the similarity between flat bands with nonzero Chern numbers
and Landau levels realized in a two-dimensional electron gas, it was
proposed that these lattice models could display a fractional quantum Hall 
effect for partially filled bands 
if electron-electron interaction is taking into account 
\cite{neupert2011fractional,sun2011nearly,tang2011high}.
Indeed, numerical evidences for the stability of fractional Chern insulator
phases were later reported \cite{sheng11,regnault11}.
We should mention that, recently, the infinite density matrix
renormalization group (iDMRG) technique was employ to study the
stability of fractional quantum Hall states in  
correlated Hofstadter-like models \cite{andrews20,andrews21}.

Correlation effects in a spinfull topological Hubbard model on a
square lattice with nearly-flat noninteracting bands but at a
commensurate filling were also discussed
\cite{neupert2012topological, doretto2015flat,su2019ferromagnetism}.
Here the noninteracting limit of the topological Hubbard model is
given by the $\pi$-flux model, whose parameters can be adjusted such
that the band structure is given by two lower and two higher
(doubly degenerated) nearly-flat bands separated by an energy gap
\cite{neupert2011fractional}.  
At $1/4$-filling (half-filling of the lower band), it was shown
\cite{neupert2012topological, doretto2015flat,su2019ferromagnetism} 
that such topological Hubbard model can realize a flat-band
ferromagnetic phase \cite{flat-fm}.
In particular, one of us calculated the spin-wave excitation spectrum
of the flat-band ferromagnetic phase within a bosonization formalism 
\cite{doretto2015flat}. For the corresponding correlated Chern
insulator, it was shown that the spin-wave spectrum has one gapped 
excitation branch and one gapless one, with the Goldstone mode at the
centre of the first Brillouin zone. These analytical results
qualitatively agrees with the numerical ones determined via exact
diagonalization \cite{su2019ferromagnetism}.

\begin{figure}[t]
\centerline{ \includegraphics[width=6.0cm]{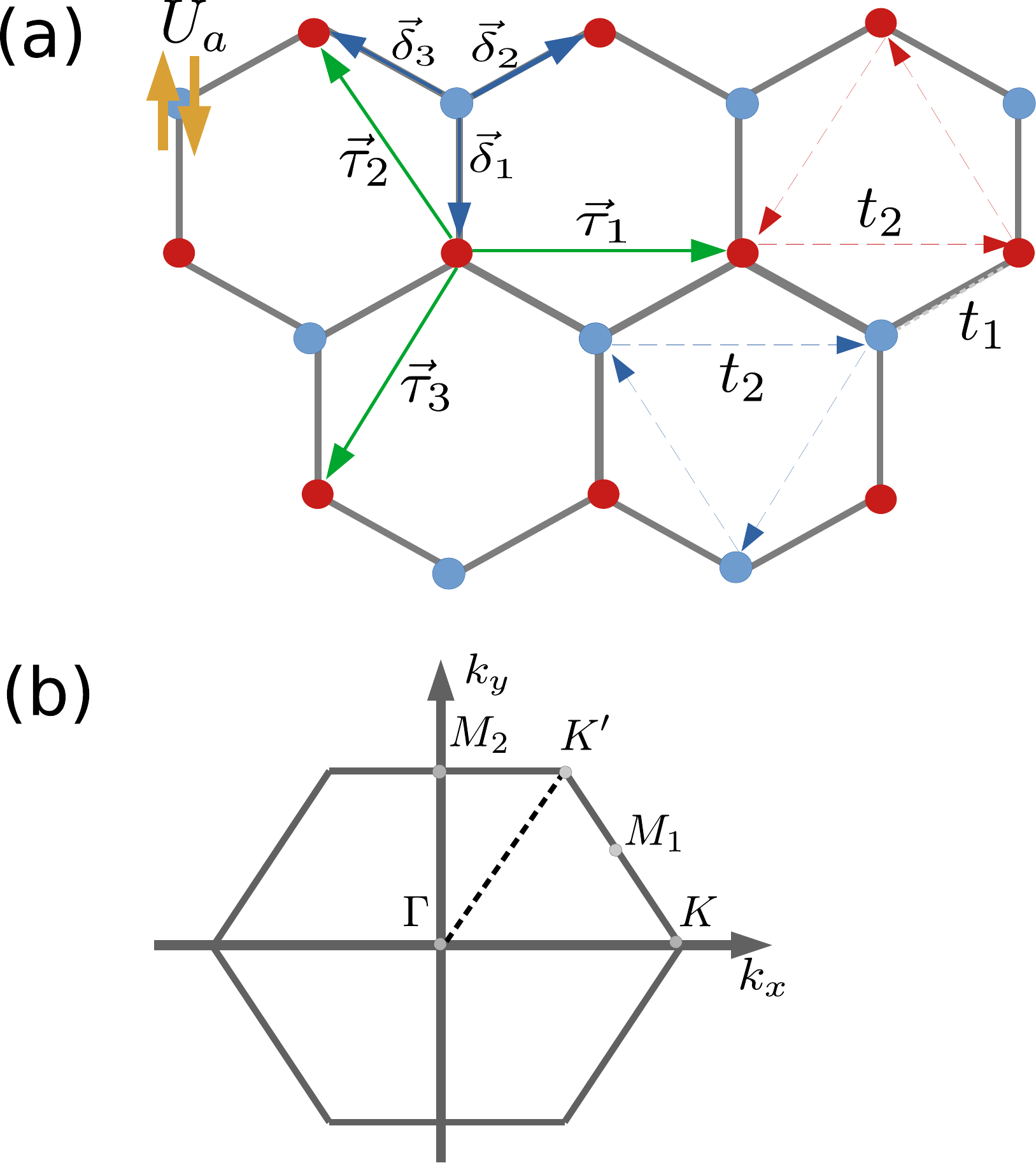} }
\caption{(a) Schematic representation of the Haldane-Hubbard model
\eqref{eqHH0} on the honeycomb lattice, indicating the
nearest-neighbor $t_1$ and next-nearest-neighbor 
$t_2 e^{\pm i \phi}$ hoppings and the on-site Hubbard
repulsion energy $U_a$. 
Blue and red circles indicate the sites of the (triangular)
sublattices $A$ and $B$, respectively.
$\deltab_i$ and $\taub_i$ are the nearest-neighbor \eqref{deltavectors}
and next-nearest-neighbor \eqref{tauvectors} vectors, respectively.
(b) The first Brillouin zone and it's highly symmetrical  points:
$\mathbf{K} = ( 4\pi/3\sqrt{3}, 0 )$,
$\mathbf{K'} = ( 2\pi/3\sqrt{3}, 2\pi/3 )$,
$\mathbf{M}_1 = ( \pi/\sqrt{3}, \pi/3)$, and
$\mathbf{M}_2 = ( 0, 2\pi/3 )$.
The nearest-neighbor distance of the honeycomb lattice $a = 1$.
}
\label{figLattice}
\end{figure}

In the present paper, we study the flat-band ferromagnetic phase of a
correlated Chern insulator on a honeycomb lattice described by the
Haldane-Hubbard model. We consider configurations close to the
nearly-flat band limit of the lower (noninteracting) bands that was previously
determined for the (spinless) Haldane model \cite{neupert2011fractional}.
We describe the flat-band ferromagnetic phase of the Haldane-Hubbard
model within the bosonization formalism for flat-band correlated Chern
insulators proposed in Ref.~\cite{doretto2015flat}. 
Such a bosonization scheme is based on the method proposed to study
the quantum Hall ferromagnetic phase of a two-dimensional electron
gas at filling factor $\nu=1$ \cite{doretto2005lowest}. It was later
employed to described the quantum Hall ferromagnetic phases realized
in graphene at filling factors $\nu = 0$ and $\nu = \pm 1$
\cite{doretto2007quantum}.  
We show that the bosonization scheme allow us to map the
Haldane-Hubbard model at the nearly-flat band limit of its lower band
to an effective interacting boson model. 
Our main finding is the flat-band ferromagnetic phase spin-wave
spectrum, which corresponds to the dispersion relation of 
the bosons determined from the  effective boson model within a
harmonic approximation. We find that the spin-wave excitation 
spectrum has one gapped and one gapless excitation branches, with a
Goldstone mode at the center of the first Brillouin zone and Dirac
points at the $K$ and $K'$ points [see Fig.~\ref{figEspectro1}(a), below]. 
Introducing an energy offset in the on-site Hubbard repulsion energies
associated with the (triangular) sublattices $A$ and $B$, one finds
that an energy gap opens in the spin-wave excitation spectrum at 
the $K$ and $K'$ points.
The effects on the spin-wave spectrum due to the presence of a
staggered on-site energy term related to the sublattices $A$ and $B$
is also discussed.

Our paper is organized as follows. 
In Sec.~\ref{sec:TBmodel}, we introduce the Haldane-Hubbard model on a
honeycomb lattice and discuss in details the band structure of the
noninteracting term close to the nearly flat-band limit determined in
Ref.~\cite{neupert2011fractional}.  
In Sec.~\ref{sec:boso}, we briefly review the bosonization scheme for
flat-band Chern insulators \cite{doretto2015flat}.  
Sec.~\ref{sec:flatferromagnetism} is devoted to the description of the
flat-band ferromagnetic phase of the Haldane-Hubbard model: 
we quote the expression of the effective interacting boson model 
derived within the bosonization scheme and determine
the spin-wave excitation spectra in the nearly flat-band limit and
slightly away from this limit;
the effects of an energy offset in the on-site Hubbard repulsion term
and of a staggered on-site energy term are also discussed.   
In Sec.~\ref{sec:summary}, 
we discuss our results and provide a brief summary of our main
findings. 
Details of the bosonization formalism are presented in Appendices 
\ref{ap:functions} and \ref{ap:BosoDetails} while additional results
derived within the bosonization scheme for the topological Hubbard
model on a square lattice previously studied  in Ref.~\cite{doretto2015flat}
are reported in Appendix~\ref{ap:square}.

\section{Haldane-Hubbard model}
\label{sec:TBmodel}

Let us consider $N_e$ spin-$1/2$ electrons on a honeycomb lattice 
described by the Haldane-Hubbard model, whose Hamiltonian is
given by 
\cite{he2011chiral,he2011topological,maciejko2013topological,
hickey15,hickey2016haldane,zheng15,vanhala16,wu2016quantum,arun16,troyer16}.  
\begin{equation}
    H =  H_0 + H_U, 
\label{eqHH0}
\end{equation}
where 
\begin{align}
 H_0 &=  t_1 \sum_{i \in A, \delta, \sigma}  \left( c_{i A \sigma}^{\dagger} c_{i + \delta B \sigma}  
                      + {\rm H.c.} \right) 
\nonumber \\
	&+ t_2  \sum_{i \in A, \tau, \sigma}   
                 \left( e^{-i\phi} c_{i A \sigma}^{\dagger} c_{i + \tau  A \sigma}   + {\rm H.c.} \right) 
\nonumber \\
        &+ t_2  \sum_{i \in B, \tau, \sigma}   
                 \left( e^{+i\phi} c_{i B \sigma}^{\dagger} c_{i + \tau  B \sigma}   + {\rm H.c.} \right)
\label{eqHH2}
\end{align}
and
\begin{equation}
   H_U = \sum_i \sum_{a = A,B} U_a \hat{\rho}_{i a \uparrow} \hat{\rho}_{i a \downarrow}.
\label{eqHH}
\end{equation}
Here the operator $c_{i a \sigma}^{\dagger}$ ($c_{i a \sigma}$) creates (destroys) 
an electron with spin $\sigma = \uparrow, \downarrow$ on site $i$ 
of the (triangular) sublattice $a=A$, $B$ of the honeycomb lattice.   
$t_1 \ge  0$ and $t_2 e^{\pm i \phi}$ with $t_2 \ge 0$ are, respectively, the
nearest-neighbor and next-nearest-neighbor hoppings.
One notices that the electron acquires a $+\phi$ ($-\phi$) phase as it moves
in the same (opposite) direction of the arrows within the same
sublattice [see dashed lines in Fig.~\ref{figLattice}(a)].
Indeed, the complex next-nearest-neighbor hopping $t_2 e^{\pm i \phi}$
results in a fictitious flux pattern with zero net flux per unit cell  
[see, e.g,  Fig.~1(a) from Ref.~\cite{neupert2011fractional} for details].  
The index $\delta$ corresponds to the nearest-neighbor vectors
[Fig.~\ref{figLattice}(a)]
\begin{align}
 \deltab_1 &= -a\hat{y}, 
\nonumber \\
 \deltab_2 &= \frac{a}{2}\left( \sqrt{3}\hat{x} + \hat{y} \right), \quad\quad
 \deltab_3 = -\frac{a}{2}\left( \sqrt{3}\hat{x} - \hat{y} \right),
\label{deltavectors}
\end{align}
while $\tau$ indicates the next-nearest-neighbor vectors
\begin{align}
 \taub_1 &= \deltab_2 - \deltab_3 = a\sqrt{3}\hat{x}, 
\nonumber \\
 \taub_2 &= \deltab_3 - \deltab_1 = -\frac{a}{2}\left( \sqrt{3}\hat{x} - 3\hat{y} \right), 
\nonumber \\
 \taub_3 &= \deltab_1 - \deltab_2 = -\frac{a}{2}\left( \sqrt{3}\hat{x} + 3\hat{y} \right).
\label{tauvectors}
\end{align}
In the following, we set the nearest-neigbhor distance to unit, i.e.,
$a = 1$.  
Finally, the $H_U$ term [Eq.~\eqref{eqHH}] is the one-site Hubbard repulsion term,
which represents an energy cost paid by double occupation of site $i$, 
and 
\begin{equation}
   \hat{\rho}_{i a\sigma} = c_{i a \sigma}^{\dagger} c_{i a \sigma}
\label{dens-op}
\end{equation}
is the density operator associated with electrons with spin $\sigma$
at site $i$ of sublattice $a$. We consider that 
the one-site repulsion energy $U_a > 0$ can depend on the sublattice $a$.

\subsection{Tight-binding term with nearly-flat topological bands}
\label{sec:FreeModel}

In this section, we discuss in details the noninteracting term $H_0$
[Eq.~\eqref{eqHH2}] of the Hamiltonian \eqref{eqHH0} and show that it
can display an almost flat (lower) electronic band that is topologically
nontrivial.

\begin{figure*}[t]
\centerline{
  \includegraphics[width=5.4cm]{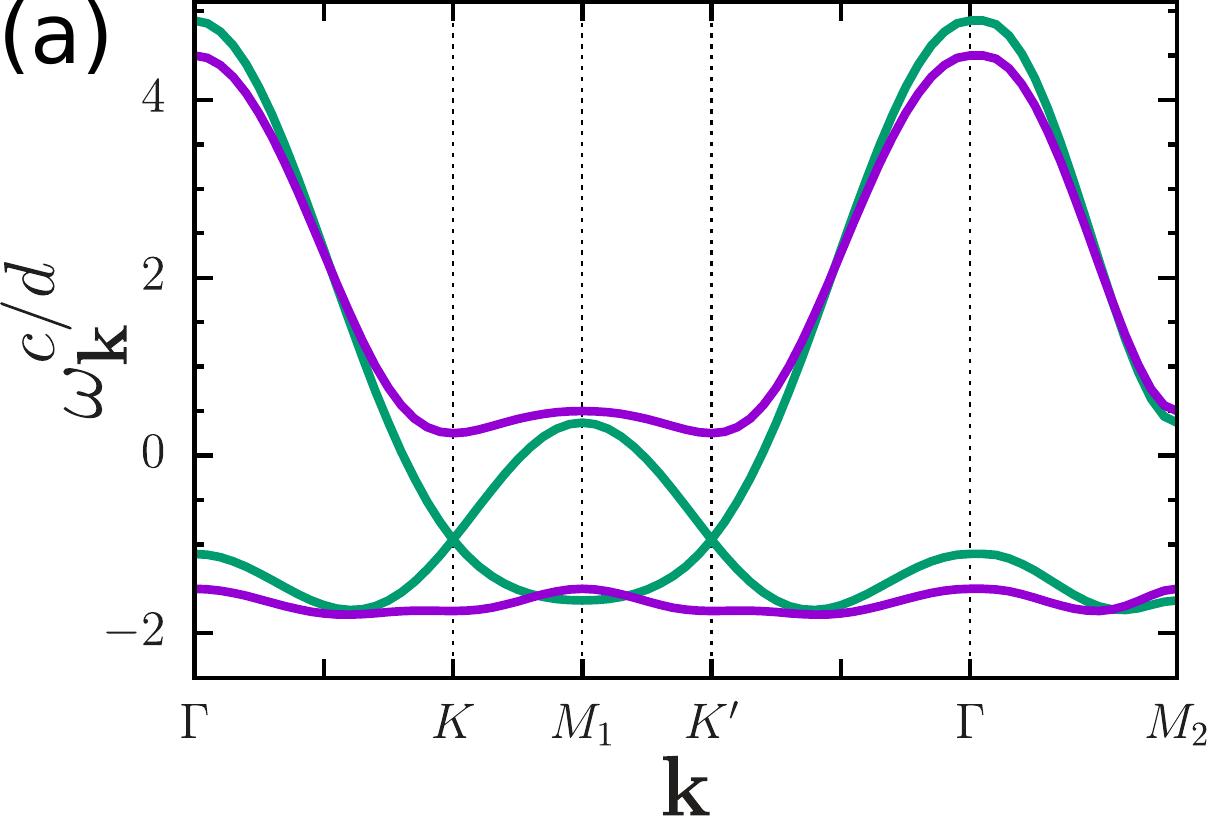}
\hskip0.7cm
  \includegraphics[width=5.4cm]{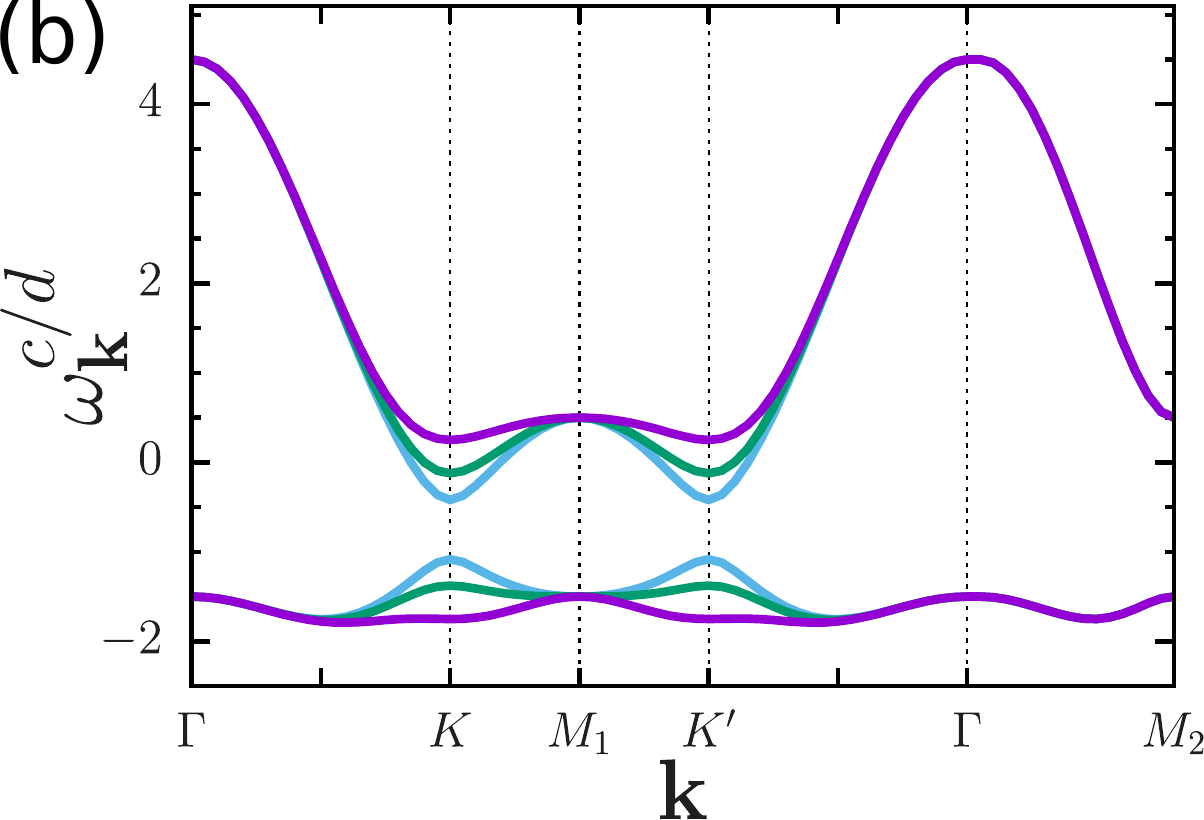}
\hskip0.7cm
  \includegraphics[width=5.4cm]{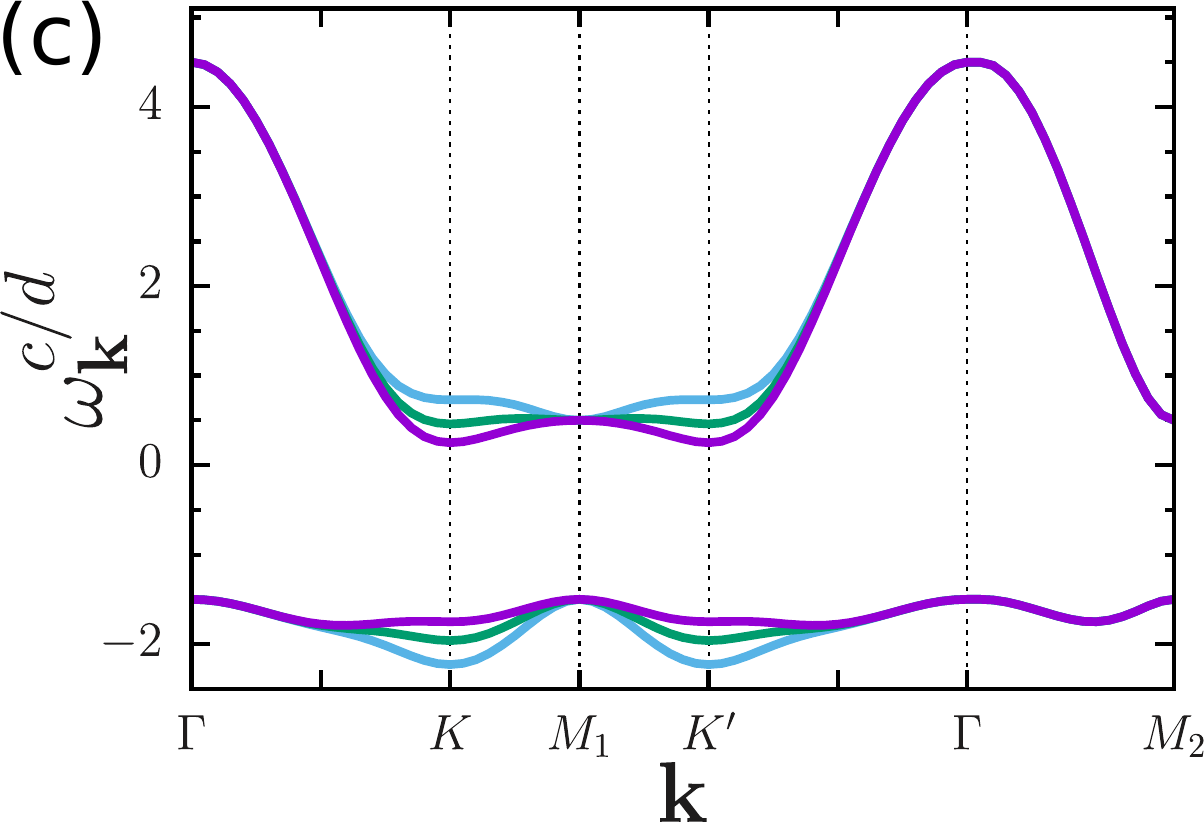}
} 
\caption{ Band structure \eqref{eq:omega} of the noninteracting hopping term
 \eqref{eqHH2} (in units of the nearest-neighbor hopping amplitude $t_1$) 
 along paths in the first Brillouin zone [Fig.~\ref{figLattice}(b)] for
 different values of the next-nearest-neighbor hopping amplitude $t_2$ 
 and phase $\phi$:
 (a) $t_2 = 0.3155\, t_1$, $\phi=0$ (green) and    
      $t_2 = 0.3155\, t_1$, $\phi=0.656$ (magenta);   
 (b) $\phi = 0.4$ (blue), 
      $\phi = 0.5$ (green), and 
      $\phi = 0.656$ (magneta), 
      with $t_2$ given by the relation $\cos(\phi) = t_1/ (4 t_2)$; and
 (c) $\phi = 0.656$ (magneta) 
      $\phi = 0.75$ (green), and 
      $\phi = 0.85$ (blue),  
      with $t_2$ given by the relation $\cos(\phi) = t_1/ (4 t_2)$.
} 
\label{figEspectro}
\end{figure*}

The first step to diagonalize the free-electron Hamiltonian \eqref{eqHH2} 
is to perform a Fourier transform,
\begin{equation}
  c_{i a \sigma}^{\dagger} = \frac{1}{\sqrt{N_a}} \sum_{\bk \in {\rm BZ}} 
                  e^{-i \mathbf{k} \cdot \mathbf{R}_i} c_{ \mathbf{k} \, a \, \sigma}^{\dagger} ,
\label{eq:Fourier}
\end{equation}
where the momentum sum runs over the first Brillouin zone (BZ)
[Fig.~\ref{figLattice}(b)] associated with the underline triangular
Bravais lattice and $N_a = N$ is the number of sites of the sublattice $a$.
It is then easy to show that the noninteracting Hamiltonian \eqref{eqHH2} 
can be written in a matrix form, 
\begin{equation}
  H_0 = \sum_{\mathbf{k}}  \Psi_{\mathbf{k} }^{\dagger}  H_\bk \Psi_{\mathbf{k} },  
\label{eqH0k}
\end{equation}	
where the $4 \times 4$ $H_\bk$ matrix reads
\begin{equation}
  H_\bk =  \left(\begin{array}{cc}
			h^{\uparrow}_{\mathbf{k}} &  0 \\ 
			0  &  h^{\downarrow}_{\mathbf{k}}
			\end{array}  \right)
\end{equation}
and the four-component spinor $\Psi_{\mathbf{k}} $ is given by 
\begin{equation}
\Psi_{\mathbf{k}} = \left( 
			c_{ \mathbf{k} A \uparrow} \;\; 
			c_{ \mathbf{k} B \uparrow} \;\;
			c_{ \mathbf{k} A \downarrow} \;\; 
			c_{ \mathbf{k} B \downarrow} \right)^T.
\end{equation}
The $2 \times 2$ matrices $h^\sigma_\bk$ associated with each spin
sector are such that $h^\uparrow_\bk  = h^\downarrow_\bk = h_\bk$, with the $h_\bk$
matrix given by 
\begin{align}
h_\bk  &= \left( \begin{array}{cc}
         2 t_2\sum_{\tau} \cos(\mathbf{k} \cdot \taub + \phi) 
         & t_1 \sum_{\delta} e^{i \mathbf{k} \cdot \deltab } \\ 
	    t_1 \sum_{\delta} e^{-i \mathbf{k} \cdot \deltab } 
     & 2 t_2 \sum_{\tau} \cos(\mathbf{k} \cdot \taub - \phi )
				   \end{array}  \right).
\nonumber
\end{align}
It is possible to write the $h_\bk$ matrix in terms of the identity
matrix $\tau_0$  and the vector $\hat{\tau} = (\tau_1$, $\tau_2$ ,$\tau_3)$, whose
components are Pauli matrices, 
\begin{equation}
    h_\bk = B_{0, \mathbf{k} } \tau_0 +  \mathbf{B}_{\mathbf{k} } \cdot \hat{\tau} ,
\label{eqPauli}
\end{equation}	
where the $B_{0, \mathbf{k} }$ function and the components of the vector  
$\mathbf{B}_{\mathbf{k} }=( B_{1,\mathbf{k} }, B_{2, \mathbf{k} }, B_{3, \mathbf{k} })$ 
are given by 
\begin{align}
  B_{0,\mathbf{k} } & = 2 t_2 \cos(\phi )\sum_{\mathbf{\tau}} \cos(\mathbf{k} \cdot \taub ) , 
\nonumber \\
  B_{1,\mathbf{k} } &=  t_1  \sum_{\mathbf{\delta}} \cos(\mathbf{k} \cdot \deltab) ,
\nonumber \\ 
  B_{2,\mathbf{k} } &=  t_1  \sum_{\mathbf{\delta}} \sin(\mathbf{k} \cdot \deltab ) ,
\nonumber \\  
  B_{3,\mathbf{k} } &=  -2 t_2 \sin(\phi ) \sum_{\mathbf{\tau}} \sin(\mathbf{k} \cdot \taub ),
\label{eqBs}	
\end{align}
with the indices $\delta$ and $\tau$ corresponding to the
nearest-neighbor \eqref{deltavectors} and next-nearest-neighbor 
\eqref{tauvectors} vectors, respectively. 
The fact that the matrices $h^\sigma_\bk$ related to each spin sector 
$h^\uparrow_\bk  = h^\downarrow_\bk = h_\bk$ indicates that the
noninteracting model \eqref{eqHH2} breaks time-reversal symmetry
(see, e.g, Appendix A from Ref.~\cite{doretto2015flat} for details).

The Hamiltonian \eqref{eqH0k} can be diagonalized via the canonical
transformation    
\begin{align}
  d_{ \mathbf{k} \sigma} = u_{\mathbf{k}} c_{ \mathbf{k} A \sigma} + v_{\mathbf{k}} c_{ \mathbf{k} B \sigma} ,
\nonumber \\
  c_{ \mathbf{k} \sigma} = v_{\mathbf{k}}^{*} c_{ \mathbf{k} A \sigma} - u_{\mathbf{k}}^{*} c_{ \mathbf{k} B \sigma},
\label{eq:BogoTransf}
\end{align}
where the coefficients $u_\bk$ and $v_\bk$ are given by 
\begin{align}	
 |u_{\mathbf{k}}|^2 &= \frac{1}{2} \left(1+\hat{B}_{3, \mathbf{k}} \right),  
\quad 
 |v_{\mathbf{k}}|^2  =   \frac{1}{2} \left(1-\hat{B}_{3, \mathbf{k}} \right),
\nonumber \\
 u_{\mathbf{k}} v_{\mathbf{k}}^{*} &=  \frac{1}{2} \left( \hat{B}_{1, \mathbf{k}} + i \hat{B}_{2, \mathbf{k}} \right),
\label{eq:Bogocoef} 
\end{align}
with the hatted $B_i$ standing for the $i$-component of the normalized
vector $\hat{\mathbf{B}}_\bk = \mathbf{B}_\bk /|\mathbf{B}_\bk|$. 
After the diagonalization, the Hamiltonian \eqref{eqH0k} then reads 
\begin{align}
  H_0 = \sum_{\mathbf{k} \sigma } 
              \omega_{\mathbf{k}}^c c_{\mathbf{k} \sigma}^{\dagger}  c_{\mathbf{k} \sigma}
           + \omega_{\mathbf{k}}^d d_{\mathbf{k} \sigma}^{\dagger}  d_{\mathbf{k} \sigma},
\label{eq:Hfree}
\end{align}
with the dispersions of the lower band $c$ ($-$ sign)  
and the upper one $d$ ($+$ sign) given by
\begin{align}
    \omega^{d/c}_{\mathbf{k}} = B_0 \pm \sqrt{B_{1, \mathbf{k}}^2 + B_{2, \mathbf{k}}^2 + B_{3, \mathbf{k}}^2 } .
\label{eq:omega}
\end{align}
Notice that both $c$ and $d$ free-electronic bands are doubly degenerated
with respect to the spin degree of freedom. 

Figure \ref{figEspectro}(a) shows the electronic bands \eqref{eq:omega}
along paths in the first Brillouin zone [Fig.~\ref{figLattice}(b)] for two different
parameter sets. 
For $t_2 = 0.3155\, t_1$ and $\phi=0$,  the spectrum is gapless due to 
the presence of Dirac points at the $K$ and $K'$ points, i.e., the
upper and lower bands touch at these points and the bands disperse
linearly with momentum around them.
A finite phase $\phi$ breaks time-reversal symmetry and opens a
gap $\Delta$ between the lower and upper bands at the $K$ and $K'$
points, as exemplified for the
parameter choice $t_2 = 0.3155\, t_1$ and $\phi=0.656$.   
Moreover, a finite phase $\phi$ yields free-electronic bands
topologically nontrivial, since the corresponding Chern numbers 
\cite{kane13, neupert2011fractional} 
\begin{equation}
 C_{\sigma}^{c/d} =\pm \frac{1}{4 \pi} \int_{BZ} d^2k 
     \hat{\mathbf{B}}_k \cdot (\partial_{k_x} \hat{\mathbf{B}}_k \times \partial_{k_y} \hat{\mathbf{B}}_k )  
\label{eqCn}
\end{equation}
are finite. One finds that $C^c_\sigma = +1$ and $C^d_\sigma = -1$ 
respectively for the lower and the upper bands regardless the spin. 
As mentioned in the Introduction,
such nonzero Chern numbers combined with broken time-reversal
symmetry indicates that the gapped phase of the noninteracting model
\eqref{eqHH2} at half-filling is indeed a Chern band insulator 
\cite{kane13,rpp-rachel18}.
The phase diagram $t_2/t_1$ v.s. $\phi$ for the noninteracting model
\eqref{eqHH2} at half-filling can be found, e.g., in Ref.~\cite{hickey15}: 
in addition to a (gapped) Chern band insulator phase with quantized
Hall conductivity $\sigma_{xy} = \pm e^2/h$ per spin, the model 
also displays a Chern metal phase with nonquantized $\sigma_{xy}$.

For the parameter choice (nearly flat-band limit)
\begin{equation}
  t_2 = 0.3155 \, t_1 \quad\quad {\rm and} 
  \quad\quad \phi=0.656, 
\label{optimal-par}
\end{equation}
one also sees that the lower band $c$ is almost flat. Indeed, such a choice obeys
the relation $\cos(\phi) = t_1/ (4 t_2) = 3\sqrt{3/43}$ 
\cite{neupert2011fractional} which yields a large flatness ratio 
for the lower band $f_c = \Delta/W_c = 6$, where 
$\Delta = {\rm min}(\omega_{d,\bk}) - {\rm max}(\omega_{c,\bk})$
is the energy gap and 
$W_c = {\rm max}(\omega^c_\bk) - {\rm min}(\omega^c_\bk)$
is the width of the lower band $c$. 
It is easy to see that the flatness ratio decreases as one moves away
from the optimal parameter choice \eqref{optimal-par}. 
For instance, in Fig.~\ref{figEspectro}(b), we plot the band structure
\eqref{eq:omega} for $\phi = 0.4$, $0.5$, and $0.656$ with $t_2$ given by  
$\cos(\phi) = t_1/ (4 t_2)$. One notices that, as the phase $\phi$
decreases, the flatness ratio $f_c$ also decreases, i.e., the energy
gap at the $K$ and $K'$ points decreases while the band width $W_c$ of
the lower band $c$ increases.
The flatness ratio $f_c$ also decreases for $\phi > 0.656$, see
Fig.~\ref{figEspectro}(c).

In the following, we consider configurations close to the
nearly flat-band limit \eqref{optimal-par} of the lower band $c$.
It is worth mentioning that previous studies 
\cite{he2011chiral,he2011topological,zheng15,arun16,troyer16,vanhala16}  
about the Haldane-Hubbard model \eqref{eqHH0} focus on configurations
with $\phi=\pi/2$, which yields a particle-hole symmetric
noninteracting band structure.

\begin{figure}[b]
\centerline{
  \includegraphics[width=6.5cm]{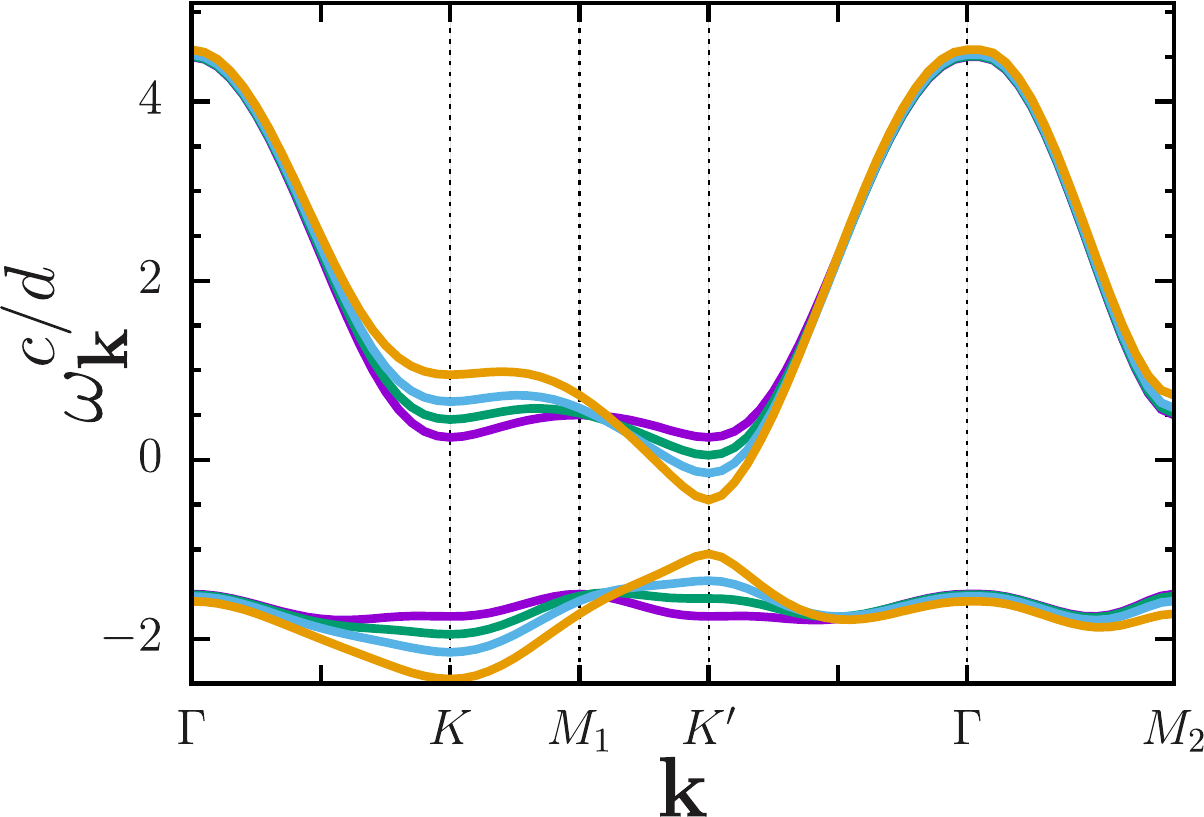}
} 
\caption{Band structure \eqref{eq:omega} of the noninteracting hopping term
 \eqref{eqHH2} with the additional staggered on-site energy term \eqref{eq:Hmass}
 (in units of the nearest-neighbor hopping amplitude $t_1$) 
 along paths in the first Brillouin zone for the next-nearest-neighbor
 hopping amplitude $t_2 = 0.3155\, t_1$, phase $\phi=0.656$,
 and staggered on-site energy
 $M =0$ (magenta), 
 $0.1 $ (green), 
 $0.2 $ (blue), and 
 $0.3\, t_1$ (orange).
  } 
\label{figEspectro2}
\end{figure}

\subsection{Staggered on-site energy term}
\label{sec:smass}

An additional interesting term, that is also present in Haldane's
original model \cite{haldane1988model}, is a staggered on-site energy term that
breaks inversion-symmetry:  
\begin{equation}
H_M = M \sum_{i \sigma} \left( c_{i A \sigma}^{\dagger} c_{i A \sigma} 
                                         -  c_{i B \sigma}^{\dagger} c_{i B \sigma} \right).
\label{eq:Hmass}
\end{equation}	
For $|M/t_2| < 3\sqrt{3}|\sin\phi|$, it was shown that the
electronic bands are topologically nontrivial \cite{haldane1988model}. 
Later, considering a topological Uhlmann number to characterize
symmetry-protected topological phases at finite temperatures,
it was verified that such a topological phase is stable up to 
some critical temperature $T_c$ \cite{delgado14}.

Adding the staggered on-site energy term \eqref{eq:Hmass} to the
tight-binding model \eqref{eqHH2}, one easily finds that the new
Hamiltonian $H_0 + H_M$ also assumes the form \eqref{eqH0k} with the
$B_{0,\bk}$ and the $B_{i,\bk}$ ($i=1,2,3$) functions given by
Eq.~\eqref{eqBs} apart from the replacement
\begin{equation}
   B_{3, \mathbf{k}} \rightarrow B_{3, \mathbf{k}} + M.
\label{B3M}
\end{equation}
In Fig.~\ref{figEspectro2}, we plot the band structure 
\eqref{eq:omega} for the parameters \eqref{optimal-par} and 
$M = 0$, $0.1$, $0.2$, and $0.3\, t_1$.
We notice that, for a finite on-site energy $M > 0$ ($M < 0$), the
energy gap is located at the $K'$ ($K$) point. 
Moreover, as the parameter $M$ increases, 
the energy gap decreases,
the difference ($\omega^c_{K'} - \omega^c_K$) increases, and
the flatness ratio of the lower band $c$ decreases. 
Indeed, the increasing of the parameter $M$ can induce a gap closure
that destroys the topological phase, 
see Fig.~2 from Ref.~\cite{haldane1988model}.

In Sec.~\ref{sec:dispersiviness} below, we consider a finite staggered
on-site energy $M$ as a source of departure of the lower band $c$ from the
nearly flat-band limit \eqref{optimal-par}.

\subsection{Hubbard term in momentum space}
\label{sec:hubbard}

The expression of the on-site Hubbard interaction \eqref{eqHH2} easily
follows from the Fourier transform of the electron density operator
\eqref{dens-op}, which is given by
\begin{equation}
  \hat{\rho}_{i a \sigma} = \frac{1}{N} \sum_{\bq \in {\rm BZ}} 
                  e^{i \mathbf{q} \cdot \mathbf{R}_i} \hat{\rho}_{a \sigma}(\bq).
\label{eq:Fourier-rho}
\end{equation}
Substituting Eq.~\eqref{eq:Fourier-rho} into the Hamiltonian
\eqref{eqHH2}, one finds
\begin{equation}
   H_U = \frac{1}{N}\sum_{a = A,B} \sum_\bq U_a 
              \hat{\rho}_{a \uparrow}(-\bq) \hat{\rho}_{a \downarrow}(\bq).
\label{hu-k}
\end{equation}

It is also useful to determine the expansion of the density operator 
$\hat{\rho}_{a \sigma}(\bk)$ in terms of the fermion operators $c_{\bk\,a\,\sigma}$.
With the aid of Eqs.~\eqref{dens-op}, \eqref{eq:Fourier}, and \eqref{eq:Fourier-rho},
one shows that
\begin{equation}
   \hat{\rho}_{a \sigma}(\bq) = \sum_\bp c^\dagger_{\bp-\bq\, a\,\sigma}c_{\bp\, a\,\sigma}.
\label{density-op-2}
\end{equation}
The canonical transformation \eqref{eq:BogoTransf} allows us to
express \eqref{density-op-2} in terms of the fermion operators 
$c_{\bk\,\sigma}$ and $d_{\bk\,\sigma}$. In particular, the density
operator \eqref{density-op-2} {\sl projected} into the lower
noninteracting band $c$ reads \cite{doretto2015flat} 
\begin{equation}
  \bar{\rho}_{a\, \sigma}(\bq) = \sum_\bp 
      G_a(\bp,\bq)c^\dagger_{\bp-\bq\,\sigma}c_{\bp\,\sigma}, 
\label{proj-dens-op} 
\end{equation}
with the $ G_a(\bp,\bq)$ function given by Eq.~\eqref{eq:Ga}.

Once the expression of the projected density operator
\eqref{proj-dens-op} is known, 
we can determine the projection $\bar{H}_U$ of the on-site Hubbard
term \eqref{hu-k} into the noninteracting lower bands $c$. Indeed,  
$\bar{H}_U$ assumes the form \eqref{hu-k} with the replacement 
$\hat{\rho}_{a \sigma}(\bq) \rightarrow \bar{\rho}_{a \sigma}(\bq)$, i.e.,
\begin{equation}
   \bar{H}_U = \frac{1}{N}\sum_{a = A,B} \sum_\bq U_a 
              \bar{\rho}_{a \uparrow}(-\bq) \bar{\rho}_{a \downarrow}(\bq).
\label{hu-k-bar}
\end{equation}

\section{Bosonization formalism for flat-band Chern insulators}
\label{sec:boso}

In this section, we briefly summarize the bosonization scheme for a Chern
insulator introduced in Ref.~\cite{doretto2015flat} for the
description of the flat-band ferromagnetic phase of a correlated Chern
insulator on a square lattice.

\begin{figure}[t]
\centerline{
  \includegraphics[width=6.0cm]{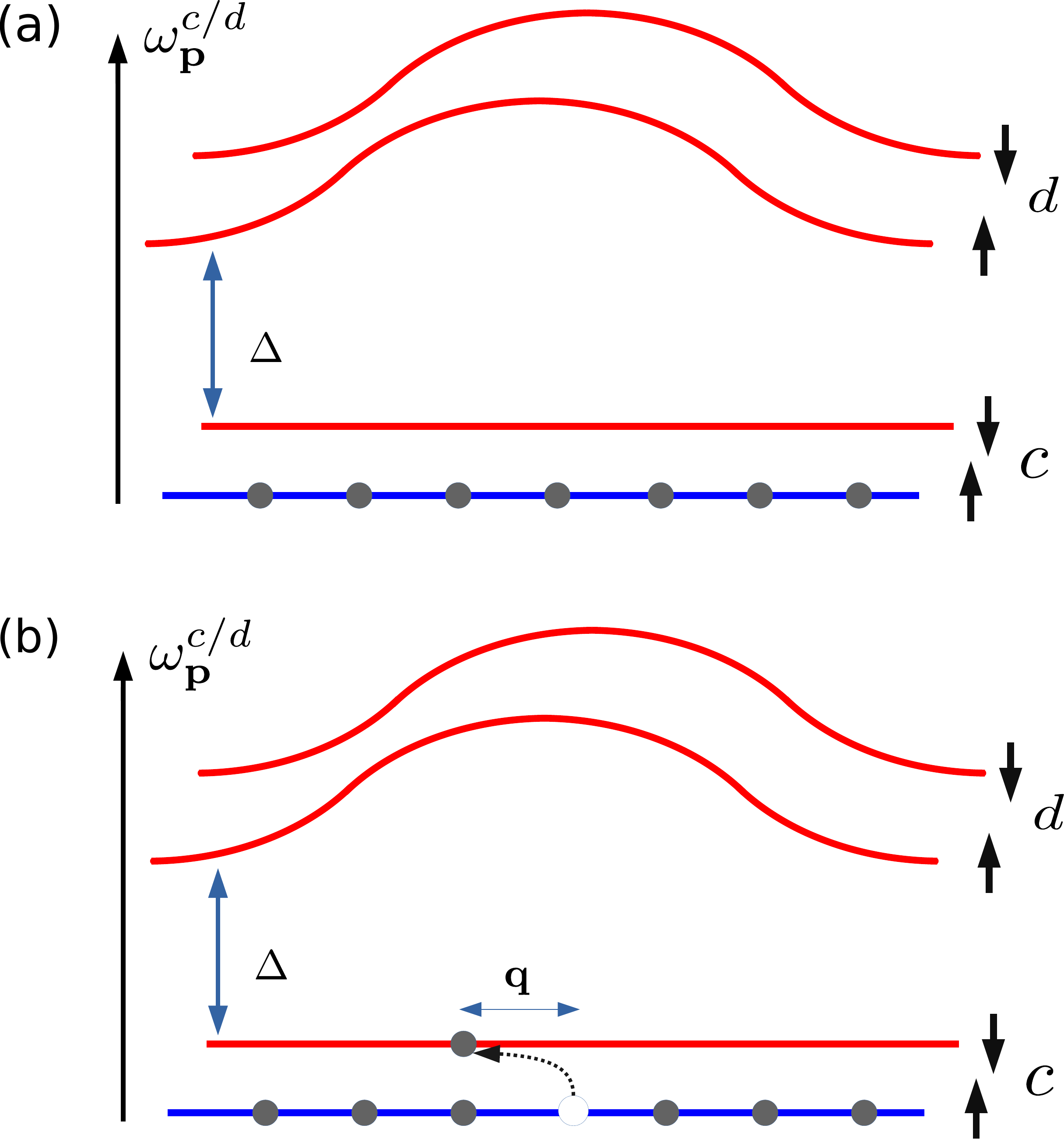}
}
\caption{Schematic representation of 
 (a) the ground state \eqref{eq:FM} of the noninteracting term \eqref{eqHH2}
 in the nearly-flat limit \eqref{optimal-par} of the lower band $c$
 at $1/4$-filling and 
 (b) the particle-hole pair excitation above the ground state \eqref{eq:FM}.
 Although the free bands $c$ and $d$ are doubly degenerated with respect
 to the spin degree of freedom, an offset between the 
 $\sigma = \uparrow$ and $\downarrow$ bands are introduced for clarity.}	
\label{figExcitation}
\end{figure}

Let us consider a spinfull Chern insulator on a bipartite lattice whose
Hamiltonian assumes the form \eqref{eqH0k}. We choose the model
parameters such that (at least) the lower band $c$ is (nearly-)
flat and consider that the number of electrons 
$N_e = N_A = N_B = N$, where $N_A$ and $N_B$ are, respectively, the
number of sites of the sublattices $A$ and $B$. Such a choice
corresponds to a $1/4$-filling, i.e., the lower (nearly-flat) band $c$
is half-filled. 
In particular, let us assume that the lower band $c \, \uparrow$ 
is completely occupied, as illustrated in Fig.~\ref{figExcitation}(a).   
In this case, the ground state of the noninteracting system 
(the {\sl reference} state) is completely spin polarized and it can be
written as a product of single particle states,  
\begin{equation}
 |{\rm FM} \rangle =  \prod_{\mathbf{k} \in BZ} c_{\mathbf{k} \uparrow}^{\dagger} |0 \rangle .
\label{eq:FM}
\end{equation}
Since the lower flat bands $c$ are separated from the upper
bands $d$ by an energy gap, the lowest-energy neutral excitations
above the ground state \eqref{eq:FM} are given by particle-hole pairs
within the lower bands $c$, i.e., they are spin-flips that can be
written as [see Fig.~\ref{figExcitation}(b)]
\begin{equation}
  | \Psi_{\mathbf{k}} \rangle \propto S_\bk^{-} | {\rm FM} \rangle.
\label{eq:Sexcitation}
\end{equation}
It is possible to show that such particle-hole pair excitations can be
treated approximated as bosons. Indeed, one can define the 
boson operators
\begin{align}
  b_{\alpha,\mathbf{q}} &= \frac{\bar{S}_{-\mathbf{q},\alpha}^{+}}{F_{\alpha\alpha,\mathbf{q}}} 
                = \frac{1}{F_{\alpha\alpha,\mathbf{q}}}  \sum_{\mathbf{p}} g_{\alpha} (\mathbf{p}, -\mathbf{q}) 
                   c_{\mathbf{p+q}\uparrow}^{\dagger} c_{\mathbf{p}\downarrow},
\nonumber \\
   b_{\alpha,\mathbf{q}}^{\dagger} &= \frac{\bar{S}_{\mathbf{q},\alpha}^{-}}{F_{\alpha\alpha,\mathbf{q}}} 
             =  \frac{1}{F_{\alpha\alpha,\mathbf{q}}}  \sum_{\mathbf{p}} g_{\alpha} (\mathbf{p}, \mathbf{q}) 
                 c_{\mathbf{p-q}\downarrow}^{\dagger} c_{\mathbf{p}\uparrow}, 
\label{eq:bosons}
\end{align}
with $\alpha = 0,1$, that satisfy the canonical commutation relations 
\begin{align}
 [b_{\alpha,\mathbf{k}} , b_{\beta, \mathbf{q}}^{\dagger}  ] &= \delta_{\alpha, \beta} \; \delta_{\mathbf{k}, \mathbf{q}}, 
\nonumber  \\ 
 [b_{\alpha,\mathbf{k}} , b_{\beta,\mathbf{q}}  ] &=  [b_{\alpha,\mathbf{k}}^{\dagger} , b_{\beta,\mathbf{q}}^{\dagger}  ] = 0, 
\label{eq:BComutations}
\end{align}
and whose vacuum state is given by the (reference) spin-polarized
state \eqref{eq:FM}, i.e.,
\begin{equation}
   b_{\alpha,\bq} | {\rm FM} \rangle = 0.
\label{vacuum} 
\end{equation}
Here the operators $\bar{S}_{\bq,\alpha}^\pm$ are linear combinations
of {\sl projected} spin operators associated with sublattices $A$ and $B$,  
\begin{equation}
   \bar{S}_{\bq,\alpha}^\pm =   \bar{S}_{\bq,A}^\pm + (-1)^\alpha \bar{S}_{\bq,B}^\pm,
\label{eq:SprojAlpha}
\end{equation}
with $\alpha = 0,1$ and $\bar{S}^\pm_{\bq,a} = \bar{S}^x_{\bq,a} \pm i \bar{S}^y_{\bq,a}$.  
The operator $\bar{S}^\lambda_{\bq,a}$, with $\lambda = x,y,z$, is the
$\lambda$-component of the spin operator $S^\lambda_{\bq,a}$ {\sl projected} 
into the lower bands $c$ with $S^\lambda_{\bq,a}$ being the Fourier
transform of the spin operator $S^\lambda_{i,a}$ at site $i$ of the
sublattice $a$.
Indeed, the projected operator $\bar{S}^\lambda_{\bq,a}$ is determined
from $S^\lambda_{i,a}$ following the same procedure outlined in
Sec.~\ref{sec:hubbard} for the density operator \eqref{proj-dens-op}.
Finally, the $F_{\alpha\beta,\mathbf{q}}$ function is given by 
\begin{equation}
 F_{\alpha \beta,\mathbf{q}}^2 = \sum_\bp \: g_{\alpha}(\mathbf{p}, \mathbf{q}) 
          g^*_{\beta}(\mathbf{p},\mathbf{q}),
\label{eq:F2}
\end{equation}
with the $g_{\alpha}(\mathbf{p}, \mathbf{q})$ function being related
to the coefficients \eqref{eq:Bogocoef} of the canonical
transformation \eqref{eq:BogoTransf},  
\begin{equation}
  g_\alpha (\mathbf{p}, \mathbf{q}) = v_{\mathbf{p}-\mathbf{q}}^* v_{\mathbf{p}} 
                                   + (-1)^{\alpha} u_{\mathbf{p}-\mathbf{q}}^* u_{\mathbf{p}}.
\label{eq:ga}
\end{equation}

Any operator expanded in terms of the fermion operators $c^\dagger_{\bk\sigma}$
and $c_{\bk\sigma}$ can, in principle, be rewritten in terms of the bosons \eqref{eq:bosons}. 
In particular, the density operator \eqref{proj-dens-op}
projected into the lower bands $c$ assumes the form 
\begin{align}
  \bar{\rho}_{a \sigma}(\mathbf{k}) &= \frac{1}{2}N\delta_{\sigma, \uparrow}\delta_{\mathbf{k}, 0} 
         + \sum_{\alpha,\beta}\sum_{\mathbf{q}} \: \mathcal{G}_{\alpha \beta a \sigma}(\mathbf{k}, \mathbf{q})  
        b_{\beta,\mathbf{k}+\mathbf{q} }^{\dagger} b_{\alpha, \mathbf{q}}, 
\label{eq:rhoBoson}
\end{align}
where the $\mathcal{G}_{\alpha \beta a \sigma}(\bk,\bq)$ function is
given by Eq.~\eqref{Gcal}.
Importantly, both $F_{\alpha \beta,\mathbf{q}}^2$ and $\mathcal{G}_{\alpha \beta a \sigma}(\bk,\bq)$ 
functions can be explicitly written in terms of the coefficients \eqref{eqBs}, 
see Eqs.~\eqref{eq:ApF201} and \eqref{Gcal2}, respectively.

\section{Flat-band ferromagnetism in the Haldane-Hubbard model}
\label{sec:flatferromagnetism}

In the completely flat-band limit (band width $W_c=0$) of the lower
noninteracting bands $c$, Hund's rule yields that the ground state of
the Haldane-Hubbard model is ferromagnetic, once such bands are
half-filled \cite{note03}. 
As the amplitude $t_2$ of the next-nearest-neighbor hopping is
modified and the noninteracting bands $c$ get more dispersive,
the ferromagnetic phase might be stable up to a critical band width $W_{c,critic}$,
for fixed on-site repulsion energies $U_a$ \cite{note04}.
Indeed, recent exact diagonalization calculations \cite{gu2019itinerant} 
indicate the stability of the flat-band ferromagnetic phase for the
Haldane-Hubbard model in the vicinity of the nearly flat-band limit
\eqref{optimal-par}, see discussion below.

In this section, we study the flat-band ferromagnet phase of the
Haldane-Hubbard model \eqref{eqHH0} within the bosonization formalism
\cite{doretto2015flat} for flat-band Chern insulators. 
In particular, we concentrate on the determination of the dispersion
relation of the elementary (neutral) particle-hole pair excitations,
i.e., we calculate the spectrum of the spin-wave
excitations above the (flat-band) ferromagnetic ground state \eqref{eq:FM}. 
We show that the spin-wave spectrum has a Goldstone mode at
momentum $\bq = 0$, a feature that indicates the stability of the
flat-band ferromagnetic ground state.

\begin{figure*}[t]
\centerline{
 \includegraphics[width=6.0cm]{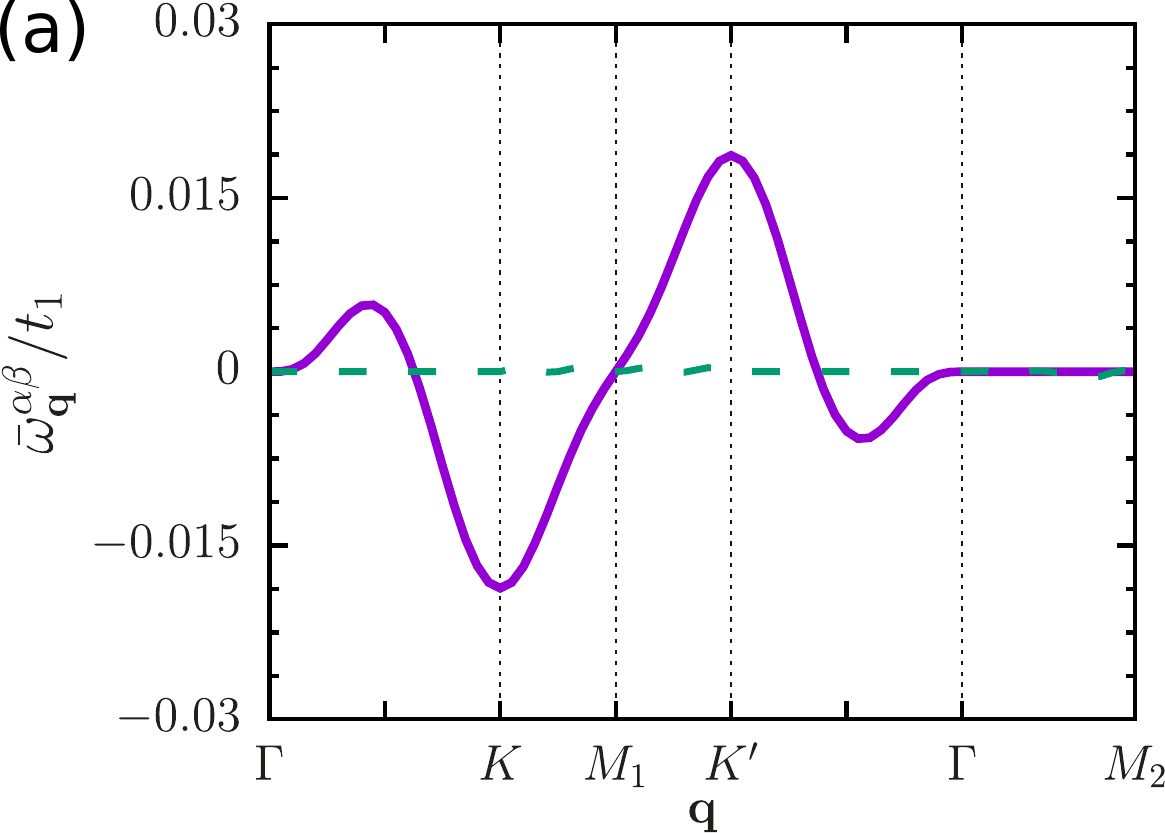}
 \hskip1.7cm
 \includegraphics[width=6.0cm]{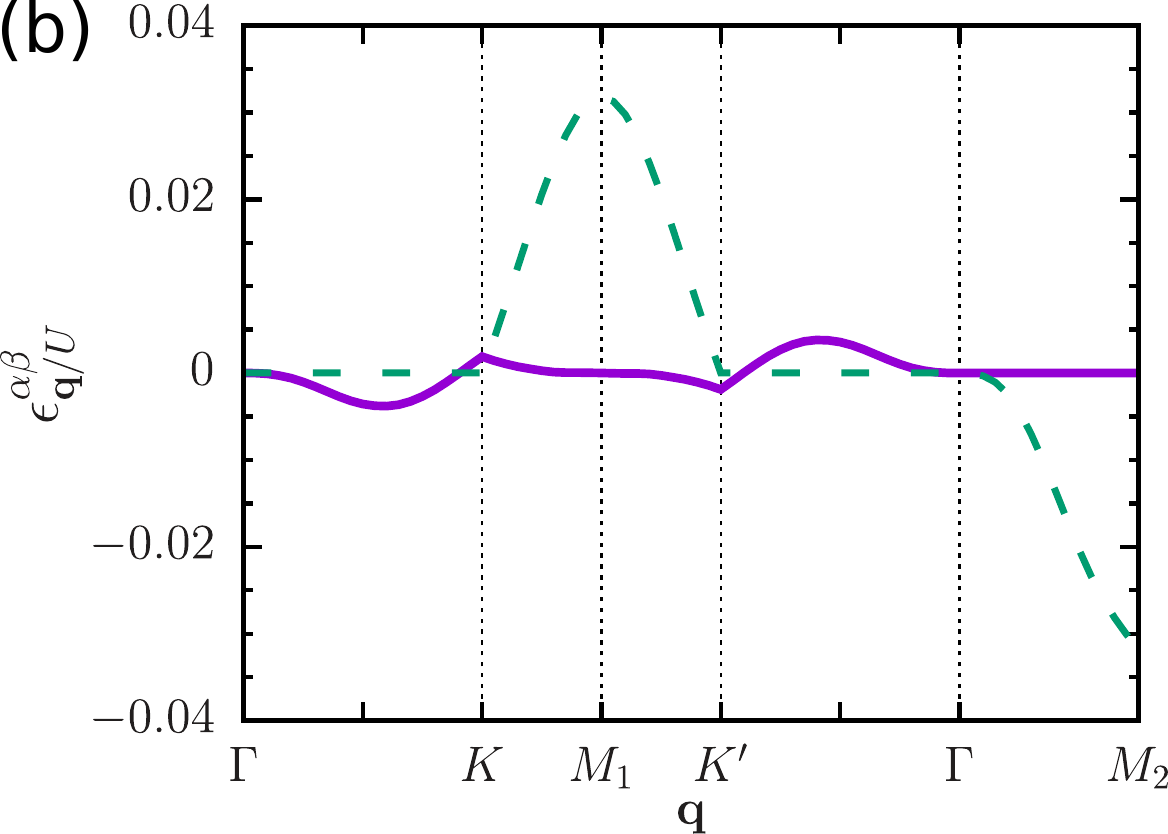}
}
\caption{The real (solid magenta line) and imaginary (dashed green line) parts of 
 (a) the coefficient $\bar{\omega}^{01}_\bq$ [Eq.~\eqref{eq:omegaBar}] and 
 (b) the coefficient $\epsilon^{01}_\bq$ [Eq.~\eqref{eq:Epsilon}]
 along paths in the first Brillouin zone [Fig.~\ref{fig:omegabar}(b)] 
 for the Haldane-Hubbard  model \eqref{eqHH0} in the nearly-flat band
 limit \eqref{optimal-par} of the lower noninteracting band $c$.}
\label{fig:omegabar}
\end{figure*}

\subsection{Effective interacting boson model}
\label{sec:ChernInsu}

Let us consider the Haldane-Hubbard model \eqref{eqHH0} 
on a honeycomb lattice with the noninteracting lower bands $c$ in the
nearly-flat band limit \eqref{optimal-par} 
and at $1/4$-filling of its corresponding noninteracting limit, i.e., 
we assume that the number of electrons 
$N_e = N_A = N_B = N$, with $N_A$ and $N_B$ being the
number of sites of the (triangular) sublattices $A$ and $B$, respectively.  
In this case, the bosonization scheme \cite{doretto2015flat} allows us
to map the Hamiltonian \eqref{eqHH0} into an effective interacting
boson model.

In order to derive such an effective boson model,
the first step is to project the Hamiltonian \eqref{eqHH0} into the
lower noninteracting bands $c$. 
Such a restriction is appropriated as long as the on-site
repulsion energies $U_a < \Delta$, where 
$\Delta = {\rm min}(\omega_{d,\bk}) - {\rm max}(\omega_{c,\bk})$
is the energy gap of the free-electronic bands (see Fig.~\ref{figEspectro}):
in particular, one finds that $\Delta = 1.75\,t_1$ for the nearly
flat-band limit \eqref{optimal-par}.
One shows that (see Eq.~(28) from Ref.~\cite{doretto2015flat} for details):  
\begin{align}
      H \rightarrow \bar{H} &= \bar{H}_0 + \bar{H}_U.
\label{H-projected}
\end{align}	
Here the projected noninteracting term $\bar{H}_0$ follows from
Eq.~\eqref{eq:Hfree}, 
\begin{align}
  \bar{H}_0 = \sum_{\mathbf{k} \sigma }  \omega^c_{ \mathbf{k}}
                     c_{\mathbf{k} \sigma}^{\dagger}  c_{\mathbf{k} \sigma},	
\label{eq:omegaProje}
\end{align}
while the projected on-site Hubbard term $\bar{H}_U$ is given by
Eq.~\eqref{hu-k-bar}. 
The expression of the noninteracting (kinetic) term $\bar{H}_0$ in terms of the
bosons \eqref{eq:bosons} is given by (see Appendix B
from Ref.~\cite{doretto2015flat} for details)
\begin{equation}
    \bar{H}_{0,B} = E_0 + \sum_{\alpha \beta} \sum_{\mathbf{q} \in BZ}  \bar{\omega}^{\alpha \beta}_{\mathbf{q} }
                           b_{\beta, \mathbf{q}}^{\dagger} b_{\alpha, \mathbf{q}},
\label{eq:H0B1}
\end{equation}
where $E_0 = \sum_\bk \omega^c_\bk$ is a constant
associated with the action of $\bar{H}_0$ into the reference state \eqref{eq:FM} and
\begin{align}
    \bar{\omega}^{\alpha \beta}_{\mathbf{q}}  &=  \frac{1}{F_{\alpha\alpha, \mathbf{q}} F_{\beta\beta, \mathbf{q}}} \sum_{\mathbf{p}}
            \left( \omega^c_{\mathbf{p-q}} - \omega^c_{\mathbf{p}} \right) 
            g_{\alpha} (\mathbf{p}, \mathbf{q})  g_{\beta}^{*} (\mathbf{p}, \mathbf{q}), 
\label{eq:omegaBar} 
\end{align}
with the $F_{\alpha\beta, \mathbf{q}}$ and the $g_{\alpha} (\mathbf{p}, \mathbf{q})$  
functions given by Eqs.~\eqref{eq:F2} and \eqref{eq:ga}, respectively.
The bosonic representation of the projected on-site Hubbard term 
$\bar{H}_U$ follows from Eqs.~\eqref{hu-k-bar} and \eqref{eq:rhoBoson}: 
After normal ordering the expression resulting from the substitution
of Eq.~\eqref{eq:rhoBoson} into \eqref{hu-k-bar}, one shows that
\cite{doretto2015flat} 
\begin{align}
     \bar{H}_{U,B}  &= \bar{H}_{U,B}^{(2)}+ \bar{H}_{U,B}^{(4)},
\end{align}
where the quadratic and quartic boson terms read 
\begin{align}
 \bar{H}_{U,B}^{(2)} &=  \sum_{\alpha \beta} \sum_{\mathbf{q} } \epsilon^{\alpha \beta}_{\mathbf{q} }
                                   b_{\beta, \mathbf{q}}^{\dagger} b_{\alpha, \mathbf{q}},
\label{H42} \\
 \bar{H}_{U,B}^{(4)} &= \frac{1}{N} \sum_{\mathbf{k} , \mathbf{q}, \mathbf{p}} \sum_{\alpha \beta \alpha' \beta'} 
                V^{\alpha \beta \alpha' \beta'}_{\mathbf{k}, \mathbf{q}, \mathbf{p} }
                        b_{\beta', \mathbf{p+k}}^{\dagger} b_{\beta, \mathbf{q-k}}^{\dagger} b_{\alpha \mathbf{q}} b_{\alpha' \mathbf{p}}, 
\label{H44}
\end{align}
with the coefficient  
\begin{align}
 \epsilon^{\alpha \beta}_{\mathbf{q} } &=  \frac{1}{2} \sum_{a}
                          U_a\mathcal{G}_{\alpha \beta a \downarrow} (0, \mathbf{q}) 
\nonumber \\
                       &+ \frac{1}{N} \sum_{a,\alpha', \mathbf{k}}
                         U_a\mathcal{G}_{\alpha' \beta a \uparrow}(-\mathbf{k}, \mathbf{k+q}) 
                           \mathcal{G}_{\alpha \alpha' a \downarrow}(\mathbf{k}, \mathbf{q})
\label{eq:Epsilon}
\end{align}
and the boson-boson interaction given by
\begin{align}
         V^{\alpha \beta \alpha' \beta'}_{\mathbf{k}, \mathbf{q}, \mathbf{p} } &= \frac{1}{N} \sum_a 
                       U_a\mathcal{G}_{\alpha \beta a \uparrow}(-\mathbf{k}, \mathbf{q}) 
                       \mathcal{G}_{\alpha' \beta' a \downarrow} (\mathbf{k}, \mathbf{p}). 
\label{eq:Vkq}
\end{align}
The effective {\sl interacting} boson model that describes
the flat-band ferromagnetic phase of the Haldane-Hubbard model \eqref{eqHH0} 
then assumes the form
\begin{align}
  \bar{H}_B = \bar{H}_{0,B} + \bar{H}^{(2)}_{U,B} + \bar{H}^{(4)}_{U,B}. 
\label{heffective}
\end{align}

\subsection{Spin-wave spectrum in the nearly flat-band limit}
\label{sec:spin-wave}

In this section, we consider the effective boson model \eqref{heffective} 
in the lowest-order (harmonic) approximation, which consists of
keeping only terms up to quadratic order in the boson
operators \eqref{eq:bosons} of the Hamiltonian \eqref{heffective}, i.e.,
we consider
\begin{align}
    \bar{H}_B  &\approx \bar{H}_{0,B} + \bar{H}^{(2)}_{U,B}.
\label{h-harm}         
\end{align}
In principle, the Hamiltonian \eqref{h-harm} can be diagonalized via a canonical
transformation yielding the spectrum of elementary excitations (spin-waves)
in terms of $\bar{\omega}^{\alpha \beta}_\bq$ [Eq.~\eqref{eq:omegaBar}] 
and $\epsilon^{\alpha \beta}_{\mathbf{q} }$ [Eq.~\eqref{eq:Epsilon}]. 
However, before proceeding, we would like to discuss both contributions in
details.

The coefficient $\bar{\omega}^{\alpha \beta}_\bq$ [Eq.~\eqref{eq:omegaBar}] 
represents the (kinetic) contribution to the energy of the elementary
excitations explicitly related to the dispersion of the 
noninteracting (lower) bands $c$. One can see that,
if the free band $c$ is completely flat ($ \omega^c_\bq =$ constant),
this coefficient vanishes while, in the nearly flat-band limit, it can be finite. 
For the noninteracting term \eqref{eqHH2} on the honeycomb lattice
in the nearly flat-band limit \eqref{optimal-par},
we find that $\bar{\omega}^{\alpha \alpha}_\bq = 0$ while
$\bar{\omega}^{01}_\bq = \bar{\omega}^{10}_\bq$ are finite but rather
small in units of the nearest-neighbor hopping energy $t_1$ 
[see Fig.~\ref{fig:omegabar}(a)].
Such a result is distinct from the square lattice $\pi$-flux model,
where symmetry considerations yield $\bar{\omega}^{\alpha \beta}_\bq = 0$
\cite{doretto2015flat}. 
We believe that the finite values of the coefficients
$\bar{\omega}^{01}_\bq$ and  $\bar{\omega}^{10}_\bq$ 
for the Haldane model
might be related not only to the symmetries of the
noninteracting Hamiltonian \eqref{eqHH2}, but also to the fact that
the condition
\begin{equation}
  F_{\alpha \beta,\bq} = \delta_{\alpha,\beta}F_{\alpha \alpha,\bq}
\label{conditionF}
\end{equation}
is not fulfilled for the Haldane model,  an important feature distinct
from the square lattice $\pi$-flux model. 
We refer the reader to Appendix \ref{ap:BosoDetails} for a detailed
discussion about the implications of the condition \eqref{conditionF} for the
approximations involved in the bosonization scheme.
Due to the smallness of $\bar{\omega}^{01}_\bq$ and  $\bar{\omega}^{10}_\bq$, 
in the following, we assume that $\bar{\omega}^{\alpha \beta}_\bq \approx 0$, 
i.e., we neglected the (explicit) kinetic contribution  
\eqref{eq:omegaBar} to the energy of the elementary excitations.

Concerning the coefficients \eqref{eq:Epsilon}, 
which are related to the one-site Hubbard term \eqref{eqHH}, we find that
$\epsilon^{\alpha \alpha}_\bq $ are real quantities while 
$\epsilon^{01}_\bq = -\epsilon^{10}_\bq = \epsilon^{10}_{-\bq} = -\epsilon^{01}_{-\bq}$ 
are complex ones, implying that the Hamiltonian \eqref{H42} is
non-Hermitian. Such a feature is also in contrast with the square
lattice $\pi$-flux model \cite{doretto2015flat} for which 
$\epsilon^{01}_\bq = \epsilon^{10}_\bq = 0$ (see also Ref.~\cite{note01}). 
In particular, for the nearly flat-band limit \eqref{optimal-par}, one
finds that $\epsilon^{01}_\bq$ is quite pronounced around the $M_1$ and $M_2$
points and it is also finite close to the $K$ and $K'$ points of the first
Brillouin zone [see Fig.~\ref{fig:omegabar}(b)]. 
Again, we believe that the non-Hermiticity of the Hamiltonian $\bar{H}^{(2)}_{U,B}$ 
might be an artefact of the bosonization scheme related to the fact that the
condition \eqref{conditionF} is not fulfilled for the Haldane model 
(see Appendix \ref{ap:BosoDetails} for details).
Since such an issue is not completely understood at the moment, in the
following, we determine the spin-wave spectrum both in the presence
and in the absence of the off-diagonal terms 
$(\alpha,\beta) = (0,1)$ and $(1,0)$ of the Hamiltonian \eqref{H42}.

The Hamiltonian \eqref{h-harm} with $\bar{\omega}^{\alpha \beta}_\bq = 0$ 
can be diagonalized via a canonical transformation similar to
Eq.~\eqref{eq:BogoTransf}, 
\begin{align}
  b_{0, \bq } = u^\dagger_\bq a_{+, \bq} + v_\bq a_{-, \bq},
\nonumber \\
  b_{1, \bq } = v^\dagger_\bq a_{+, \bq} - u_\bq a_{-, \bq},
\label{eq:BogoTransf2}
\end{align}
where the coefficients $u_\bq$ and $v_\bq$ are now given by 
\begin{align}	
 |u_\bq|^2,  |v_\bq|^2  &=   \frac{1}{2}  \pm
          \frac{1}{4\epsilon_\bq}\left( \epsilon^{00}_\bq - \epsilon^{11}_\bq \right),
\nonumber \\
 u_\bq v_\bq^{*} &=   \frac{\epsilon^{01}_\bq}{4\epsilon_\bq},
\quad
 v_\bq u_\bq^{*}  =   \frac{\epsilon^{10}_\bq}{4\epsilon_\bq},
\label{eq:Bogocoef2} 
\end{align}
with
\begin{equation}
 \epsilon_\bq = \frac{1}{2}\sqrt{ \left( \epsilon^{00}_\bq - \epsilon^{11}_\bq \right)^2
                             +  4 \epsilon^{01}_\bq \epsilon^{10}_\bq}.
\label{aux-epsilon}
\end{equation}
It is then easy to show that the Hamiltonian \eqref{h-harm} assumes
the form  
\begin{equation}
  \bar{H}_B  =  E_0 +   \sum_{\mu = \pm } \sum_{\bq \in BZ }   
   \Omega_{\mu, \bq} a_{\mu, \mathbf{q}}^{\dagger} a_{\mu, \mathbf{q}},
\label{eq:HFinalFlat}
\end{equation}
where the constant $E_0 = \sum_\bk \omega^c_\bk = (-1.69\,t_1)N$
for the nearly flat-band limit \eqref{optimal-par} and 
the dispersion relation of the bosons $a_\pm$ reads 
\begin{equation}
  \Omega_{\pm,\bq} = \frac{1}{2}\left( \epsilon^{00}_\bq + \epsilon^{11}_\bq \right) 
                                 \pm \epsilon_\bq,
\label{omega-b}
\end{equation}
with $\epsilon_\bq$ given by Eq.~\eqref{aux-epsilon} (see also Ref.~\cite{note02}).  
Assuming that $\epsilon^{01}_\bq = \epsilon^{10}_\bq = 0$, the
dispersion relation \eqref{omega-b} reduces to 
\begin{equation}
  \Omega_{-,\bq} = \epsilon^{00}_\bq
\quad \quad {\rm and} \quad\quad
  \Omega_{+,\bq} = \epsilon^{11}_\bq,
\label{omega-b2}
\end{equation}
since $\epsilon^{00}_\bq < \epsilon^{11}_\bq$.

\begin{figure}[t]
\centerline{
 \includegraphics[width=7.0cm]{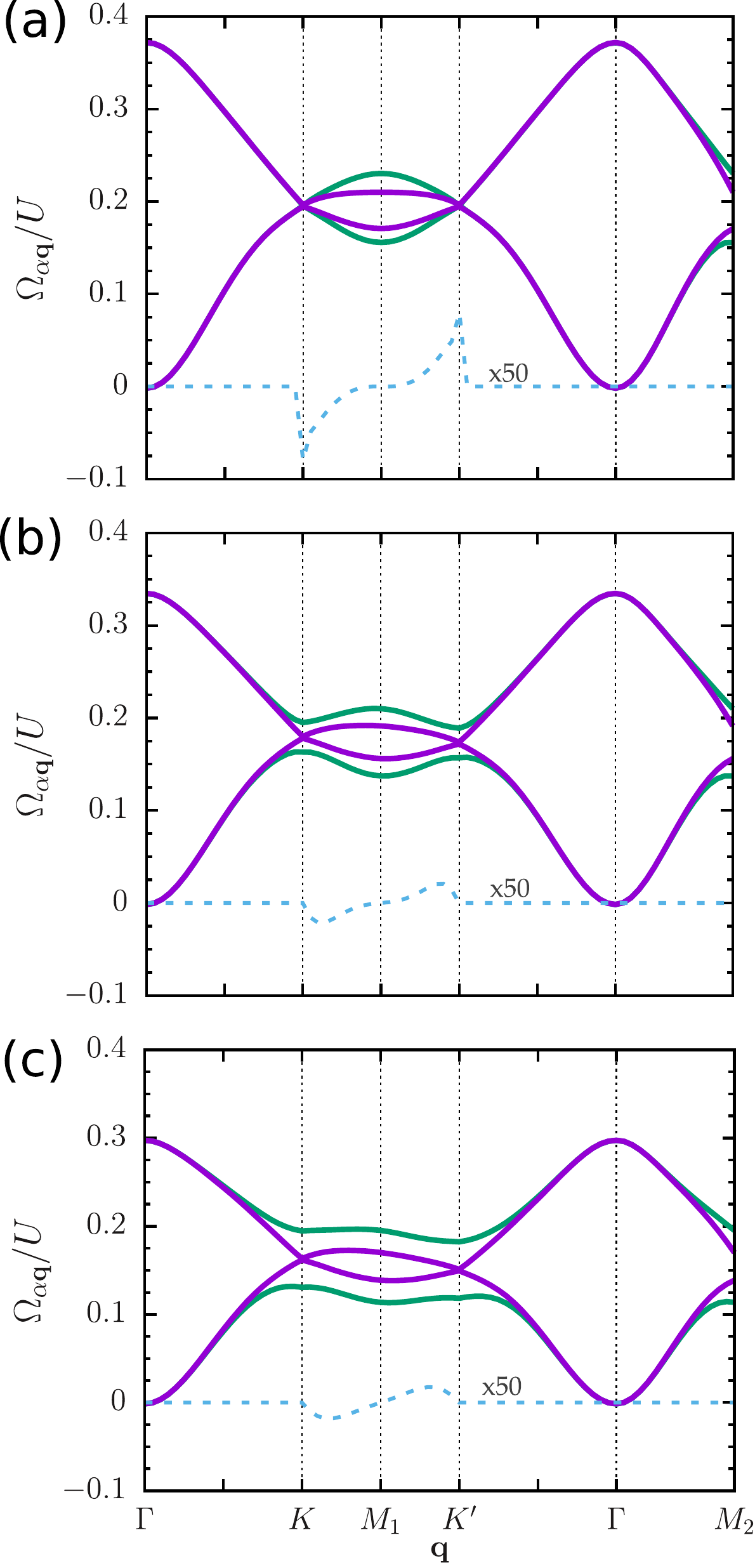} 
}
\caption{The elementary excitation (spin-wave) energies of the
  effective boson model \eqref{heffective} in the harmonic approximation for
  the nearly-flat band limit \eqref{optimal-par}:  
  dispersion relations \eqref{omega-b} (real part, solid green line) 
  and \eqref{omega-b2} (solid magenta line) along paths in the first
  Brillouin zone [Fig.~\ref{figLattice}(b)]. 
  The dashed blue line indicates the imaginary part of 
  $\Omega_{+,\bq} = -\Omega_{-,\bq} $
  [see Eq.~\eqref{omega-b}], which is multiplied by a factor of 50 for clarity.
  On-site repulsion energies:  
   (a) $U_A = U_B = U$;
   (b) $U_B = 0.8\, U_A = 0.8\, U$; and  
   (c)  $U_B = 0.6\, U_A = 0.6\, U$.}
\label{figEspectro1}
\end{figure}

One notices that the ground state of the Hamiltonian \eqref{eq:HFinalFlat} 
is the vacuum (reference) state for both bosons $b_{0,1}$ and $a_\pm$, which
corresponds to the spin-polarized ferromagnet state $|{\rm FM}\rangle$ 
[see Eqs.~\eqref{eq:FM} and \eqref{vacuum}].
Such a result is a first indication of the stability of a
flat-band ferromagnetic phase for the Haldane-Hubbard model \eqref{eqHH0}.
Indeed, one expects that the ferromagnetic ground state might be
stable only if $U \ge U_c(t_2)$. 
Unfortunately, due to limitations of our bosonization scheme, at the
moment, it is not possible to determine such critical $U_c$: 
recall that (see Sec.~\ref{sec:spin-wave}) the kinetic coefficients
\eqref{eq:omegaBar} related with the  dispersion of the noninteracting bands $c$
are not included in the effective boson model \eqref{heffective}
due to the fact that the condition \eqref{conditionF} is not fulfilled for the
Haldane model.
We refer the reader to Sec.~\ref{sec:summary} below for a more
detailed discussion about the stability of the flat-band ferromagnetic
phase.

The dispersion relations \eqref{omega-b} and \eqref{omega-b2} of the bosons $a_\pm$,
which indeed corresponds to the spin-wave spectrum above the flat-band
ferromagnetic ground state \eqref{eq:FM}, for the 
nearly flat-band limit \eqref{optimal-par}
and $U_A = U_B = U$ is shown in Fig.~\ref{figEspectro1}(a). 
Due to the absence of the kinetic coefficients \eqref{eq:omegaBar} associated with the
dispersion of the noninteracting bands $c$, one sees that the energy
scale of the spin-wave spectrum is determined by the on-site repulsion
energy $U$.
Both spin-wave spectra \eqref{omega-b} and \eqref{omega-b2} have two branches:   
the acoustic (lower) branch $\Omega_{-,\bq}$ is gapless, with a Goldstone mode
at the Brillouin zone center ($\Gamma$ point) and the 
characteristic quadratic dispersion of ferromagnetic spin-waves near
the $\Gamma$ point;
the optical (upper) one $\Omega_{+,\bq}$ is gapped, with the lowest energy
excitation at the $K$ and $K'$ points.
The presence of the Goldstone mode indicates the stability of the
flat-band ferromagnetic phase.
Interestingly, for the dispersion relation \eqref{omega-b2}, one finds  
a quite small energy gap at the $K$ and $K'$ points  
($\Delta^{(K)} = \Omega_{+,K} - \Omega_{-,K} = 2.01 \times 10^{-3}\, U$) 
while the excitation spectrum \eqref{omega-b} displays Dirac points at
the $K$ and $K'$ points. 
Indeed, the presence of the Dirac points is related to the fact that 
$\epsilon^{01}_\bq$ and $\epsilon^{10}_\bq$ are finite at the $K$
and $K'$ points, see Fig.~\ref{fig:omegabar}(b). 
Moreover, the fact that $\epsilon^{01}_\bq = -\epsilon^{10}_\bq$  
yields a very small decay rate (the imaginary part of
$\Omega_{\pm,\bq}$) for the spin-wave excitations
\eqref{omega-b} at the border of the first Brillouin zone
[see the dashed line in Fig.~\ref{figEspectro1}(a) and 
note the multiplicative factor 50].

In addition to a configuration with homogeneous on-site Hubbard energy
$U_A = U_B = U$, we also consider the Haldane-Hubbard model with a
sublattice dependent on-site Hubbard energy. 
The spin-wave spectra \eqref{omega-b} and \eqref{omega-b2} for the nearly
flat-band limit \eqref{optimal-par} and with 
$U_B = 0.8\, U_A = 0.8\, U$ and $U_B = 0.6\, U_A = 0.6\, U$ are shown 
in Figs.~\ref{figEspectro1}(b) and (c), respectively.
One notices that both spin-wave spectra \eqref{omega-b} and
\eqref{omega-b2} have a Goldstone mode at the $\Gamma$ point,
the energies of the excitations decreases as the diference 
$\Delta U = U_A - U_B$ increases, and 
the difference between the energies at the $K$ and $K'$ points  
(e.g., $\Omega_{-,K} - \Omega_{-,K'}$) also increases with $\Delta U$.
For $U_B > U_A$, we find similar features, but the energy at the $K$
point is lower than the one at the $K'$ point. 
Importantly, the dispersion relation \eqref{omega-b2} has a small
gap at the $K$ and $K'$ points, similar to the homogeneous case $U_A = U_B$:   
$\Delta^{(K)} = 1.81 \times 10^{-3}\, U$ ($\Delta U = 0.2\,U$)
and $1.61 \times 10^{-3}\, U$ ($\Delta U = 0.4\,U$).
On the other hand, for the dispersion relation \eqref{omega-b}, a
finite energy gap opens at the $K$ and $K'$ points in contrast with the 
homogeneous case $\Delta U = 0$.  One finds that
$\Delta^{(K)} = 3.18 \times 10^{-2}\, U$ ($\Delta U = 0.2\,U$)
and $6.40 \times 10^{-2}\, U$ ($\Delta U = 0.4\,U$).
Such a finite energy gap might be related to the fact that a Hubbard
term with $U_A \not= U_B$ breaks inversion symmetry.
Similar to the homogeneous configuration, the spin-wave excitations
\eqref{omega-b} at the first Brillouin zone border have a
finite decay rate.

Gu and collaborators \cite{gu2019itinerant} performed exact
diagonalization calculations and determined the spin-wave spectrum 
for the Haldane-Hubbard model \eqref{eqHH0} in the nearly flat-band
limit \eqref{optimal-par} neglecting the dispersion of the
noninteracting electronic bands, which corresponds to the
approximation $\bar{\omega}^{\alpha \beta}_\bq = 0$ considered above.   
For homogeneous on-site Hubbard energies $U_A = U_B$, it was 
found that the spin-wave spectrum has Dirac points at the $K$ and $K'$
points (see Fig.~2(a$_1$) from Ref.~\cite{gu2019itinerant})
while, for a finite $\Delta U$, the energies of the excitations
decrease with $\Delta U$ and energy gaps open at the 
$K$ and $K'$ points 
(see Figs.~2(b$_1$) and 2(c$_1$) from Ref.~\cite{gu2019itinerant}).
Remarkably, the spin-wave spectrum \eqref{omega-b} determined within
the bosonization scheme qualitatively agrees with the numerical one,  
apart from the fact that the numerical results do not indicate a
finite decay rate.

One should mention that the presence of Dirac points at the $K$ and $K'$ 
points is not only a feature of the spin-wave spectrum of the
flat-band ferromagnetic phase of the Haldane-Hubbard model.
Indeed, recent exact diagonalization calculations \cite{gu21} for a
topological Hubbard model on a kagome lattice also indicate such a
feature in the excitation spectrum of the corresponding flat-band
ferromagnetic phase when the dispersion of the (lower) noninteracting
electronic band is neglected.

As mentioned above, although the non-Hermiticity of the Hamiltonian
\eqref{H42} (and consequently finite decay rates) might be an artefact
of the bosonization scheme, the off-diagonal terms
$\epsilon^{01}_\bq$ and  $\epsilon^{10}_\bq$ of the quadratic
Hamiltonian \eqref{H42} should be considered in order to properly
describe the spin-wave spectrum at the border of the first Brillouin
zone. Therefore, in the following, we determine the spin-wave spectrum
away from the nearly flat-band limit \eqref{optimal-par} with the aid
of Eq.~\eqref{omega-b}.

\subsection{Spin-wave spectrum away from the nearly flat-band limit}
\label{sec:dispersiviness}

Although the main focus of our discussion is the description of the
flat-band ferromagnetic phase of the Haldane-Hubbard model in the
nearly-flat band limit \eqref{optimal-par}, we also consider
configurations such that the noninteracting band $c$
has smaller flatness ratio $f_c < 6$. 
In particular,  we consider the effects on the spin-wave spectrum
\eqref{omega-b} related to the increasing of the band
width $W_c$ (decreasing of the flatness ratio $f_c$) of the noninteracting
band $c$ due to 
(i) the decrease/increase of the phase $\phi$ [see
Figs.~\ref{figEspectro}(b) and (c)] and
(ii) the presence of a staggered on-site energy term \eqref{eq:Hmass} in the total
Hamiltonian (see Fig.~\ref{figEspectro2}).
These perturbations furnish some clues about the stability of the
flat-band ferromagnetic phase.

\begin{figure}[t]
\centerline{\includegraphics[width=6.5cm]{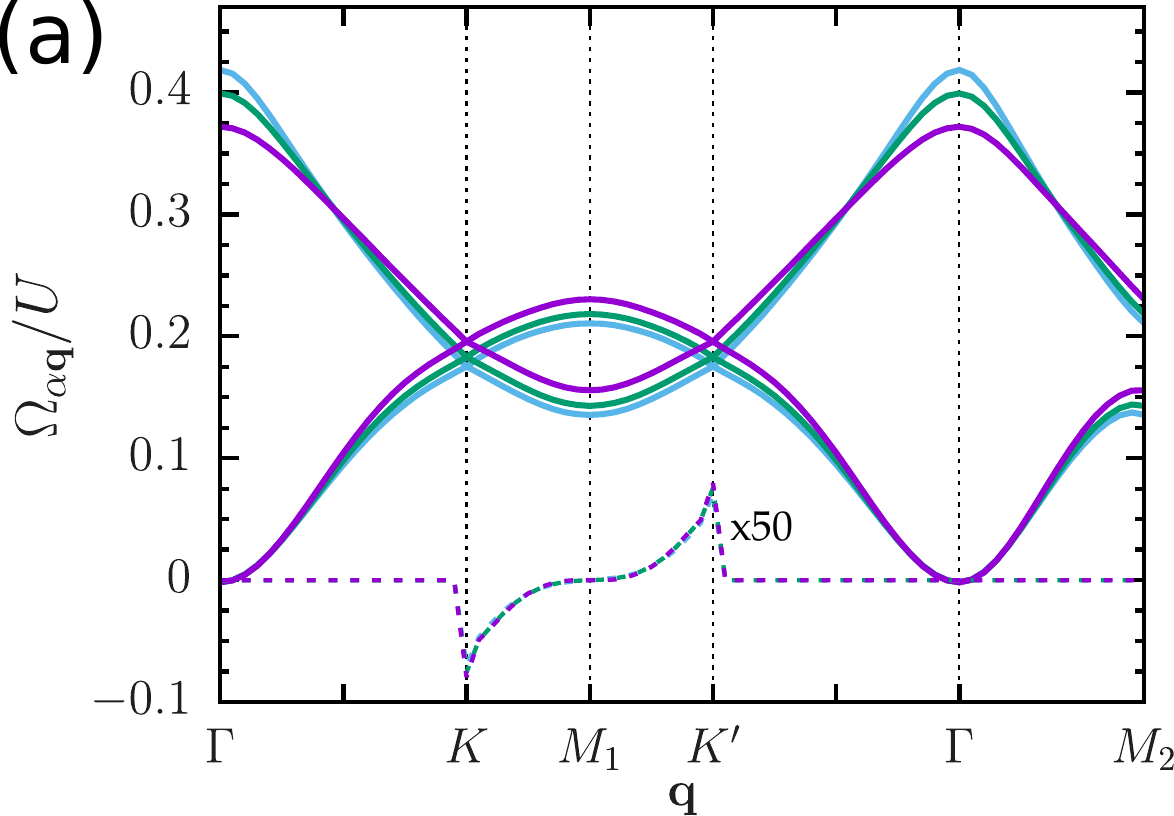}}
   \vskip0.5cm
\centerline{\includegraphics[width=6.5cm]{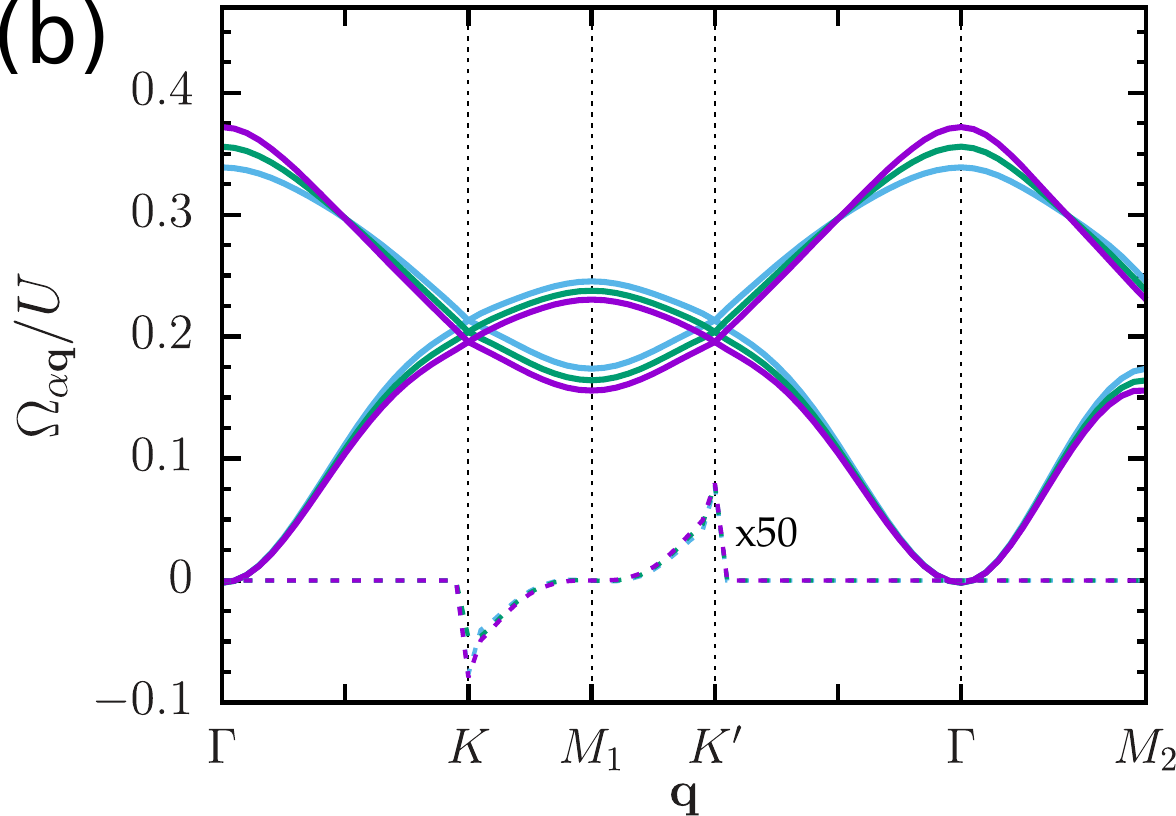}} 
\caption{The real part of the dispersion relation \eqref{omega-b}
  (solid line) along paths in the first
  Brillouin zone [Fig.~\ref{figLattice}(b)] for on-site repulsion energy 
  $U_A = U_B = U$ and
  $t_2$ given by the relation $\cos(\phi) = t_1/ (4 t_2)$.
  (a) $\phi = 0.4$ (blue), 
       $\phi = 0.5$ (green), 
       $\phi = 0.656$ (magneta) and
  (b) $\phi = 0.656$ (magneta),
       $\phi = 0.75$ (green), 
       $\phi = 0.85$ (blue).
  The corresponding dashed line indicates the imaginary part of 
  $\Omega_{+,\bq} = -\Omega_{-,\bq} $
  [see Eq.~\eqref{omega-b}], which is multiplied by a factor of 50 for clarity.}
\label{fig:Omega-phi}
\end{figure}

\begin{figure*}[t]
\centerline{
   \includegraphics[width=6.5cm]{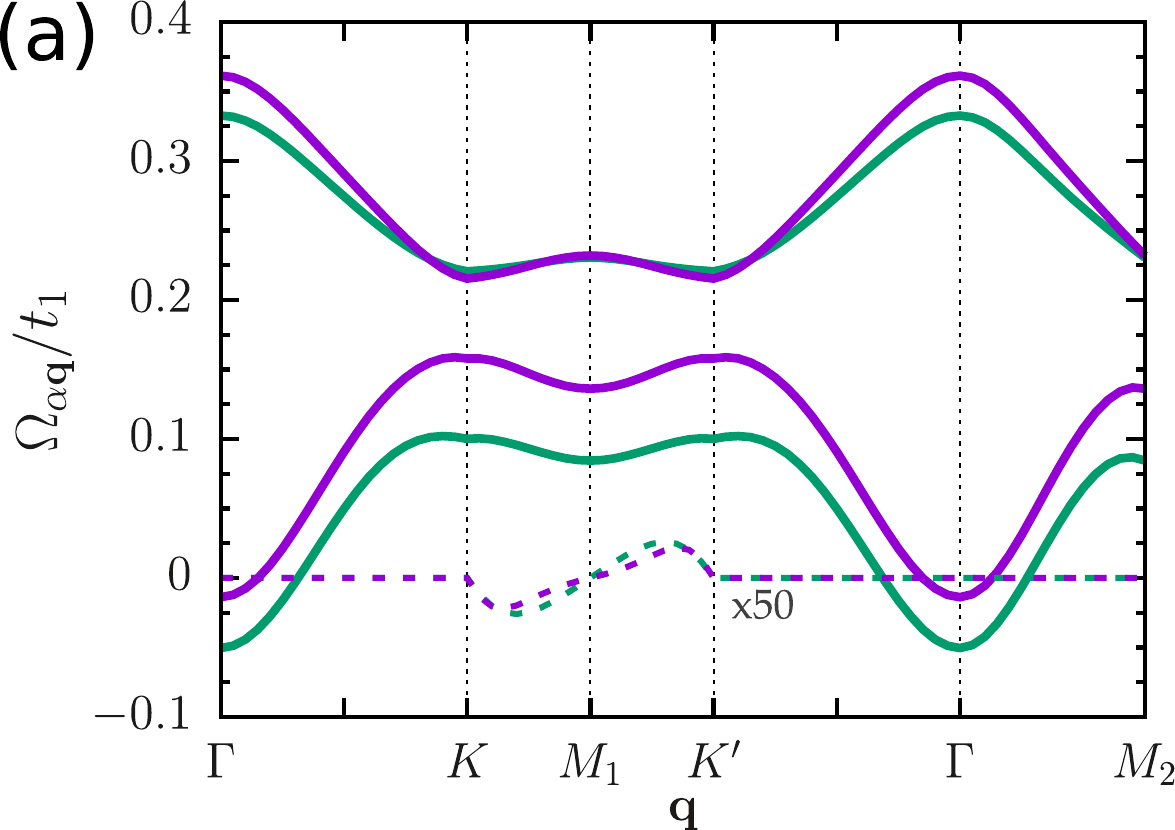}
   \hskip1.0cm 
   \includegraphics[width=6.5cm]{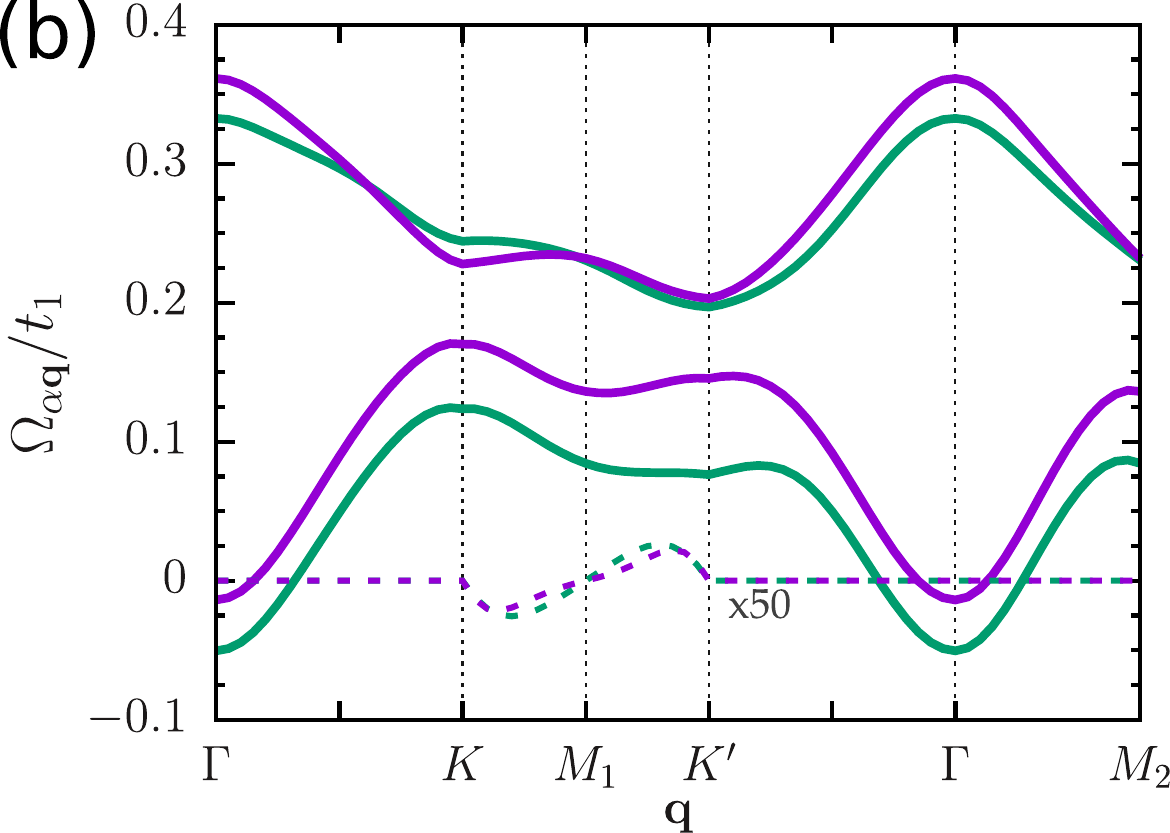}
}
\caption{The real part of the dispersion relation \eqref{omega-b}
  (solid line) along paths in the first Brillouin zone [Fig.~\ref{figLattice}(b)] 
  for the optimal parameters \eqref{optimal-par}, 
  on-site Hubbard energy $U_A = U_B = U = t_1$, 
  and staggered on-site energy 
  $M = 0.1$ (magenta) and 
  $0.2\, t_1$ (green).
  (b) Similar results for the spin-wave spectrum \eqref{omega-b},
  but with the replacement \eqref{omega-b3}. 
  The corresponding dashed line indicates the imaginary part of 
  $\Omega_{+,\bq} = -\Omega_{-,\bq} $
  [see Eq.~\eqref{omega-b}], which is multiplied by a factor of 50 for clarity.
}
\label{fig:Omega-Mass}
\end{figure*}

In Fig.~\ref{fig:Omega-phi}(a), we show the spin-wave spectrum
\eqref{omega-b} for $\phi = 0.4$, $0.5$, and $0.656$, the hopping
amplitude $t_2$ given by  
the relation $\cos(\phi) = t_1/ (4 t_2)$, and the on-site repulsion energy 
$U_A = U_B = U$. 
One sees that the spin-wave spectrum 
(in units of the on-site repulsion energy $U$) for $\phi = 0.4$ and $0.5$ is
quite similar to the one derived for the nearly-flat band limit \eqref{optimal-par}.
As the flux parameter $\phi$ decreases, the excitation energies near the
border of the Brillouin zone [the $K$-$M_1$-$K'$ line] decrease,
while the energies of the optical branch in the vicinity of the $\Gamma$ point 
increase. 
The fact that the spin-wave spectrum displays a Goldstone mode at the
$\Gamma$ point, regardless the value of the phase $\phi$, indicates the
stability of the flat-band ferromagnetic phase with respect to the simultaneous
variations of the phase $\phi$ and the next-nearest-neighbor hopping
amplitude $t_2$.    
Finite decay rates are still found at the border of the first
Brillouin zone. 
The flat-band ferromagnetic phase seems also to be stable for $\phi > 0.656$, 
see Fig.~\ref{fig:Omega-phi}(b). Here, however, 
as the flux parameter $\phi$ increases, the excitation energies near the
border of the Brillouin zone increase
and the energies of the upper branch in the vicinity of the $\Gamma$ point 
decrease.

The effect on the spin-wave spectrum of a finite staggered on-site
energy $M$ [Eq.~\eqref{eq:Hmass}] is quite distinct. 
In Fig.~\ref{fig:Omega-Mass}(a), we plot the spin-wave spectrum  
\eqref{omega-b} for the optimal parameters \eqref{optimal-par}, 
$M = 0.1$ and $0.2\, t_1$, and the on-site Hubbard energy 
$U_A = U_B = U = t_1$.  
Comparing with the homogeneous on-site energy $M = 0$ configuration 
[Fig.~\ref{figEspectro1}(a)], one sees that the whole spin-wave spectrum 
shifts downward in energy as $M$ increases and energy gaps open at the
$K$ and $K'$ points. The latter is indeed related to the fact that the
staggered on-site energy term \eqref{eq:Hmass} breaks inversion
symmetry.  
Most importantly, the energies of the acoustic branch are negative in
the vicinity of the $\Gamma$ point, indicating an instability of the
flat-band ferromagnetic phase for finite values of the staggered
on-site energy $M$. 
Such features are also found for the square lattice $\pi$-flux model
\cite{doretto2015flat}, see Fig.~\ref{fig:MassSquare} in 
Appendix \ref{ap:square}.

A finite staggered on-site energy $M$ also modifies the (kinetic) coefficients
\eqref{eq:omegaBar} directly related to the dispersion of the noninteracting
band $c$. In particular, we find that $\bar{\omega}^{\alpha\alpha}_\bq$ 
no longer vanishes for a finite $M$ [see Fig.~\ref{fig:omegabar}(a) for $M = 0$]. 
Such an effect can be easily included in the spin-wave spectrum \eqref{omega-b}
with the replacement  
\begin{equation}
   \epsilon^{\alpha \alpha}_\bq \rightarrow \bar{\omega}^{\alpha\alpha}_\bq + \epsilon^{\alpha \alpha}_\bq.
\label{omega-b3}
\end{equation}
Figure~\ref{fig:Omega-Mass}(b) shows the spin-wave spectrum \eqref{omega-b} 
with the replacement \eqref{omega-b3} 
(in units of the nearest-neighbor hopping amplitude $t_1$) 
for the optimal parameters \eqref{optimal-par},  
$M = 0.1$ and $0.2\, t_1$, and the on-site Hubbard energy 
$U_A = U_B = U = t_1$.
One notices that $\bar{\omega}^{\alpha\alpha}_\bq$ does not modify the
excitation energies in the vicinity of the $\Gamma$ point, but only
changes the excitation energies near the border $K$-$M_1$-$K'$ of the
first Brillouin zone. Such an effect resembles the one found when
distinct on-site repulsion energies $U_A \not= U_B$ are considered, see
Figs.~\ref{figEspectro1}(b) and (c).

\section{Summary and discussion}
\label{sec:summary}

The effective boson model \eqref{heffective} 
is not only restricted to the Haldane-Hubbard model \eqref{eqHH0} on a
honeycomb lattice, but, in principle, it can also be employed to study
the flat-band ferromagnetic phase of a correlated Chern insulator described by a  
topological Hubbard model on a bipartite lattice whose noninteracting
(kinetic) term breaks time-reversal symmetry and assumes the form \eqref{eqH0k}: 
Notice that a tight-binding model of the form \eqref{eqH0k} is
completely defined by the $B_{0,\bk}$ and $B_{i,\bk}$ ($i=1,2,3$)
functions \eqref{eqBs};
the $F_{\alpha\beta,\bq}$ function, which is important in the
definition of the boson operators \eqref{eq:bosons}, can be written in
terms of the normalized $\hat{B}_{i,\bk} = B_{i,\bk}/|\bB_{i,\bk}|$
functions, see Eq.~\eqref{eq:ApF201}; 
finally, the coefficients \eqref{eq:omegaBar} and \eqref{eq:Epsilon}
and the boson-boson interaction \eqref{eq:Vkq}, which completely
characterise the effective boson model \eqref{heffective}, can also be
expressed in terms of the normalized $\hat{B}_{i,\bk}$ functions, see
Appendix~\ref{ap:functions}.

An important requirement for the application of the bosonization
scheme \cite{doretto2015flat} to a topological Hubbard model is that
the condition \eqref{conditionF} is fulfilled by the noninteracting
term of the model. Once the validity of such a condition is verified,
the two sets of independent boson operators \eqref{eq:bosons} can be
defined and the bosonic representation of a operator written in terms
of the original fermions, such as the projected density operator
\eqref{proj-dens-op}, is well defined. 
This is indeed the case for the square lattice topological Hubbard
model whose noninteracting limit is giving by the $\pi$-flux model
\cite{doretto2015flat}. 
As mentioned above, the spin-wave spectrum \cite{su2019ferromagnetism} 
determined via exact diagonalization calculations for the  flat-band
ferromagnetic phase of the square lattice $\pi$-flux model in the
(completely) flat-band limit qualitatively agrees with the harmonic one
calculated within the bosonization formalism.
Additional results for the square lattice $\pi$-flux model derived
within the bosonization scheme are presented in Appendix~\ref{ap:square}.

For the Haldane-Hubbard model on a honeycomb lattice in the
nearly-flat band limit \eqref{optimal-par} of the noninteracting
(lower) band $c$ (the second application of the bosonization
formalism), the condition \eqref{conditionF} is not fulfilled for all
momenta $\bq$ [see Fig.~\ref{fig:F2}(b)],
which implies that additional considerations are needed in order
to apply the bosonization scheme.
As discussed in Appendix~\ref{ap:BosoDetails}, in order to preserve the 
form of the original effective boson model \eqref{heffective},
one should assume that the two set of boson operators $b_{0,1}$
defined by Eq.~\eqref{eq:bosons} are independent and that the bosonic
expression \eqref{eq:rhoBoson} for the projected electron density
operator $\bar{\rho}_{a\sigma}(\bk)$ holds for the Haldane model. 
Moreover, although the non-Hermiticity of the quadratic Hamiltonian
\eqref{H42} might be related to the fact that the condition
\eqref{conditionF} is not completely valid for the Haldane model, one
should keep the off-diagonal terms $\epsilon^{01}_\bq$ and $\epsilon^{10}_\bq$
[see Eq.~\eqref{eq:Epsilon}] in order to properly describe the
spin-wave excitations at the border of the first Brillouin zone
(see Fig.~\ref{figEspectro1}) as indicated by the comparison between
the dispersion relations \eqref{omega-b} and \eqref{omega-b2} and the
numerical results \cite{gu2019itinerant}.
Importantly, it is not clear at the moment whether the finite decay
rates found for high-energy spin-wave excitations are an artefact of
the bosonization scheme.
Even considering these additional approximations, the
qualitatively agreement between the real part of the dispersion
relation \eqref{omega-b} and the numerical spin-wave spectrum
\cite{gu2019itinerant} indicates that the effective boson model
\eqref{heffective} provides an appropriated description for the
flat-band ferromagnetic phase of the Haldane-Hubbard model.

In addition to the nearly-flat band limit \eqref{optimal-par}, we also
study the flat-band ferromagnetic phase of the Haldane-Hubbard model
when the noninteracting lower band $c$ gets more dispersive. 
While the ferromagnetic phase seems to be less sensible to the
increase of the band width $W_c$ of the noninteracting band $c$
due to variations of the phase $\phi$ and the next-nearest neighbor
amplitude $t_2$ (Fig.~\ref{fig:Omega-phi}), 
an instability of the ferromagnetic ground state is
found when a staggered on-site energy \eqref{eq:Hmass} is included
[Fig.~\ref{fig:Omega-Mass}(a) and (b)]. 
Interestingly, for the latter, one notices that the $F^2_{\alpha\alpha,\bq}$ 
function and ${\rm Im}\, F^2_{01,\bq} = {\rm Im}\,F^2_{10,\bq}$ are
not affected by a finite staggered on-site energy $M$, while 
${\rm Re}\, F^2_{01,\bq} = {\rm Re}\,F^2_{10,\bq}$
acquire a constant value proportional to the staggered on-site energy
$M$ [see Figs.~\ref{fig:F2}(b) and (c)]: indeed, it is easy to
see that the replacement \eqref{B3M} modifies the second term of the
integrand \eqref{eq:ApF201}, yielding an additional term proportional
to the parameter $M$. 
Notice that the instability of a flat-band ferromagnetic phase due to
a finite $M$ is also related to a stronger violation of the condition
\eqref{conditionF}. 
At the moment, it is not clear whether such an instability is an
artefact of the bosonization scheme related to some difficulties in
including kinetic effects, see the discussion below.
Interestingly, such kind of instability is also found for the square
lattice $\pi$-flux model, see Appendix~\ref{ap:square} for details.

The stability of a flat-band ferromagnetic phase was studied by
Kusakabe and Aoki via exact diagonalization calculations 
performed for the two-dimensional Mielke model \cite{kusakabe1994ferromagnetic}
and Mielke and Tasaki models \cite{kusakabe1994B}.
A parameter $\gamma$ was introduced in the original models, such that 
$\gamma = 0$ corresponds to (lower) noninteracting bands completely
flat (flat-band limit). It was found that, for $\gamma = 0$, a 
ferromagnetic phase is stable regardless the value of the on-site
repulsion energy $U$. For finite values of the parameter $\gamma$
(system away from the flat-band limit), a ferromagnetic ground state
is stable only if $U \ge U_c(\gamma)$ (see Figs.~1 and 2 from 
Ref.~\cite{kusakabe1994B}).  
Such a scenario agrees with more recently numerical results
for the square lattice $\pi$-flux model in the nearly-flat band limit
\cite{su2019ferromagnetism}, which indicates that a ferromagnetic
phase sets in only if $U \ge U_c(t_2)$, with $t_2$ being the
next-nearest neighbor hopping amplitude. 
M\"uller {\sl et al.} studied one- and two-dimensional
Hubbard models with nearly-flat bands that are not in the class of
Mielke and Tasaki flat-band models, since they do not obey some
connectivity conditions \cite{muller16}. 
They found that small and moderate noninteracting band dispersion may 
stabilize a ferromagnetic phase for $U \ge U_c$, i.e., the
ferromagnetic phase is driven by the kinetic energy.  
In particular, for a two-dimensional bilayer model,
$U_c(\delta_l)$ is a nonmonotonic function of the parameter $\delta_l$ that
controls the width of the band (see Fig.~7 from \cite{muller16}),
i.e., the ferromagnetic phase sets in only for a finite band dispersion. 
For rigorous results about the stability of a ferromagnetic phase on
Hubbard models with nearly-flat bands, we refer the reader to the 
review by Tasaki \cite{tasaki1996stability}.

The fact that a ferromagnetic phase is stable in Hubbard models with
nearly-flat (noninteracting) bands only for $U \ge U_c$ is related to
the competition between the kinetic energy (dispersion of the
noninteracting bands) and the (short-range) Coulomb interaction $U$ 
\cite{tasaki1996stability}.
The bosonization formalism \cite{doretto2015flat}
partially takes into account such a competition:
although the explicitly contribution \eqref{eq:omegaBar} of the
dispersion of the noninteracting bands $c$ is not included in the
effective boson model \eqref{heffective}, such kinetic effects are
partially considered by the bosonization scheme, since the
coefficients \eqref{eq:Epsilon}  and the boson-boson interaction
\eqref{eq:Vkq} depend on the $\hat{B}_{i,\bq}$ functions 
\eqref{eqBs} that completely determines the free-band structure
\eqref{eq:omega}.
At the moment, it is not clear how to properly include in the effective
boson model \eqref{heffective} the main effects related to the
noninteracting band dispersion.
Due to this limitation, we expected that the results derived within
the bosonization scheme for flat-band Chern insulators get more
accurate as the (lower) noninteracting bands gets less dispersive.
One should recall that the bosonization scheme \cite{doretto2015flat}
is based on the formalism \cite{doretto2005lowest} that was proposed
to describe the quantum Hall ferromagnet realized in a two-dimensional
electron gas at filling factor $\nu=1$: here, the noninteracting
bands corresponds to (completely flat) Landau levels.

Concerning the topological properties of the spin-wave excitations,
one would expect that the nontrivial topological properties of the 
noninteracting electronic bands of the Haldane-Hubbard model 
may yield a flat-band ferromagnetic phase with topologically
non-trivial spin-wave excitation bands. 
Indeed, topological magnons in Heisenberg ferromagnets 
\cite{zhang13,malki2019topological,owerre2016first,owerre2016topological,chen18,mook20}
and, in particular, magnets on a honeycomb lattice 
\cite{owerre2016first,owerre2016topological,chen18,mook20}
have been studied. An important ingredient for such topological magnon
insulators is the Dzyaloshinskii-Moriya interaction that may open
energy gaps in the magnon spectrum and yields magnon bands with 
nonzero Chern numbers. 
We calculate the Chern numbers of the spin-wave bands for
configurations of the Haldane-Hubbard model whose spin-wave spectrum  
displays an energy gap at the $K$ and $K'$ points 
[Figs.~\ref{figEspectro1}(b) and (c)]:   
we expand the Hamiltonian \eqref{h-harm} in terms of Pauli matrices as
done in Eq.~\eqref{eqPauli}, determine the corresponding $B_{i,\bq}$
coefficients assuming that $\epsilon^{01}_\bq = (\epsilon^{10}_\bq)^*$, 
and calculate the Chern numbers using Eq.~\eqref{eqCn}. 
In agreement with exact diagonalization calculations 
(see Figs.~2(b$_1$) and (c$_1$) from \cite{gu2019itinerant})
for the Haldane-Hubbard model in the nearly flat-band limit \eqref{optimal-par}
and neglecting the dispersion of the noninteracting electronic bands
(similar to the approximation $\bar{\omega}^{\alpha \beta}_\bq = 0$ 
considered in Sec.~\ref{sec:spin-wave}),
we find that the Chern numbers of the spin-wave bands vanish.
Such a result is indeed a feature of the completely flat-band limit:
the exact diagonalization calculations \cite{gu2019itinerant}
indicate that the spin-wave excitation bands have nonzero Chern
numbers only when the dispersion of the 
noninteracting electronic bands is explicitly taking into account
(see Figs.~2(a$_2$) and (d) from \cite{gu2019itinerant});
as discussed in the previous paragraph, at the moment, it is not
clear how to include in the effective boson model \eqref{heffective}
the main effects associated with the dispersion of the  noninteracting
electronic lower bands $c$.

In summary, in this paper we studied the flat-band ferromagnetic
phase of a correlated Chern insulator on a honeycomb lattice described
by the Haldane-Hubbard model. 
We considered the system at $1/4$-filling of the noninteracting bands
and in the nearly-flat band limit of the noninteracting lower bands.
We determined the spin-wave excitation spectrum within a bosonization
scheme for flat-band correlated Chern insulators and found that it has a
Goldstone mode at the first Brillouin zone center and Dirac
points at the $K$ and $K'$ points.
We also studied how the spin-wave excitation spectrum changes 
as an offset in the  on-site Hubbard energies associated with the
sublattices $A$ and $B$ is introduced
and as the width of the lower noninteracting bands increase due to 
variations of the kinetic term  parameters and 
the presence of a staggered on-site energy term. In particular, we
found that the flat-band ferromagnetic phase might be unstable when a
finite staggered on-site energy term is included in the kinetic term
of the Haldane-Hubbard model.

The bosonization scheme for flat-band correlated Chern insulators
provides an effective interacting boson model for the description of
the flat-band ferromagnetic phase of a topological Hubbard model.
In the near future, we intend to study the effects of the boson-boson
interaction not only in the Haldane-Hubbard model, but also in the
square lattice $\pi$-flux model previously studied in
Ref.~\cite{doretto2015flat}. Motivated by the similarities with the
quantum Hall ferromagnetic phase realized in a two-dimensional
electron gas at filling factor $\nu=1$ \cite{doretto2005lowest}, we
would like to check whether the boson-boson interaction may yield
two-boson bound states.

We also intend to investigate how the bosonization scheme can be
modified in order to properly include the main effects associated with the
dispersion of the noninteracting electronic bands that are encoded in the
kinetic coefficients $\bar{\omega}^{\alpha \beta}_\bq$ [Eq.~\eqref{eq:omegaBar}].
Recall that, for the square lattice topological Hubbard model
previously studied \cite{doretto2015flat}, 
symmetry considerations yield $\bar{\omega}^{\alpha \beta}_\bq=0$,
while, for the Haldane model, a (small) finite $\bar{\omega}^{\alpha \beta}_\bq$
might be related to the fact that the condition \eqref{conditionF} is
not fulfilled for the Haldane model.
Once this is done, we will be able to properly describe flat-band
ferromagnetic phases of the Haldane-Hubbard model with topologically
nontrivial spin-wave excitation bands. 
Moreover, we could identify possible instabilities of the ferromagnetic 
ground state associated with a softening of the spin-wave excitation 
spectrum at finite momentum $\bq$. 
Indeed, such a feature was observed in exact diagonalization
calculations for a square lattice topological Hubbard model when the
effects of the dispersion of the noninteracting electronic bands are
explicitly taken into account
(see Fig.~5 from Ref.~\cite{su2019ferromagnetism}).
A softening of the spin-wave bands at finite momentum $\bq$ indicates
that the Haldane-Hubbard model at $1/4$-filling may display other
magnetic ordered phases,
such as a chiral tetrahedron ordered state that was discussed in a
previous mean-field analysis of the Hubbard model on a honeycomb
lattice \cite{li12}.
Importantly: in principle, the bosonization scheme could not be
employed to study such distinct magnetic ordered phases, since the 
definition of the bosons operators \eqref{eq:bosons} is based on the
ferromagnetic (reference) state \eqref{eq:FM}.

It would be interesting to see whether the bosonization formalism
\cite{doretto2015flat}, eventually combined with the approximations
discussed in this paper, can also be employed to study twisted bilayer
graphene near a magic angle
\cite{cao2018unconventional, lu19, sharpe19, morell10, xu18, wu18, seo19, sau20,repellin20,chen20}. 
Here the resulting moir\'e pattern induces an effective 
superlattice and a set of flat-minibands in the moir\'e Brillouin zone. 
In addition to a superconducting phase \cite{cao2018unconventional, lu19}, 
evidences for a ferromagnetic phase at $3/4$-filling of the conduction
miniband are also found \cite{sharpe19}. 
In principle, a possible flat-band ferromagnetic phase of the
effective lattice model introduced in Ref.~\cite{seo19} for twisted
bilayer graphene could be studied within the bosonization scheme.
It would be interesting to compare such results with 
recent numerical ones obtained within exact diagonalization \cite{sau20,repellin20}.

Finally, we should mention that a similar study reported in this paper
for a correlated  flat-band Z$_2$ topological insulator on a honeycomb
lattice, described by a topological Hubbard model similar to Eq.~\eqref{eqHH0}
but that preserves time-reversal symmetry, is currently in progress
and it will be published elsewhere.  
For $\phi = \pi/2$, such a model corresponds to the
Kane-Mele-Hubbard model, see Ref.~\cite{rpp-rachel18} for details.

\acknowledgments

L.S.G.L. kindly acknowledges the financial support of Brazil "Ministry of Science,
Technology and Innovation” and the “National Council for Scientific
and Technological Development – CNPq”.

\appendix

\begin{figure*}[t]
\centerline{
  \includegraphics[width=5.5cm]{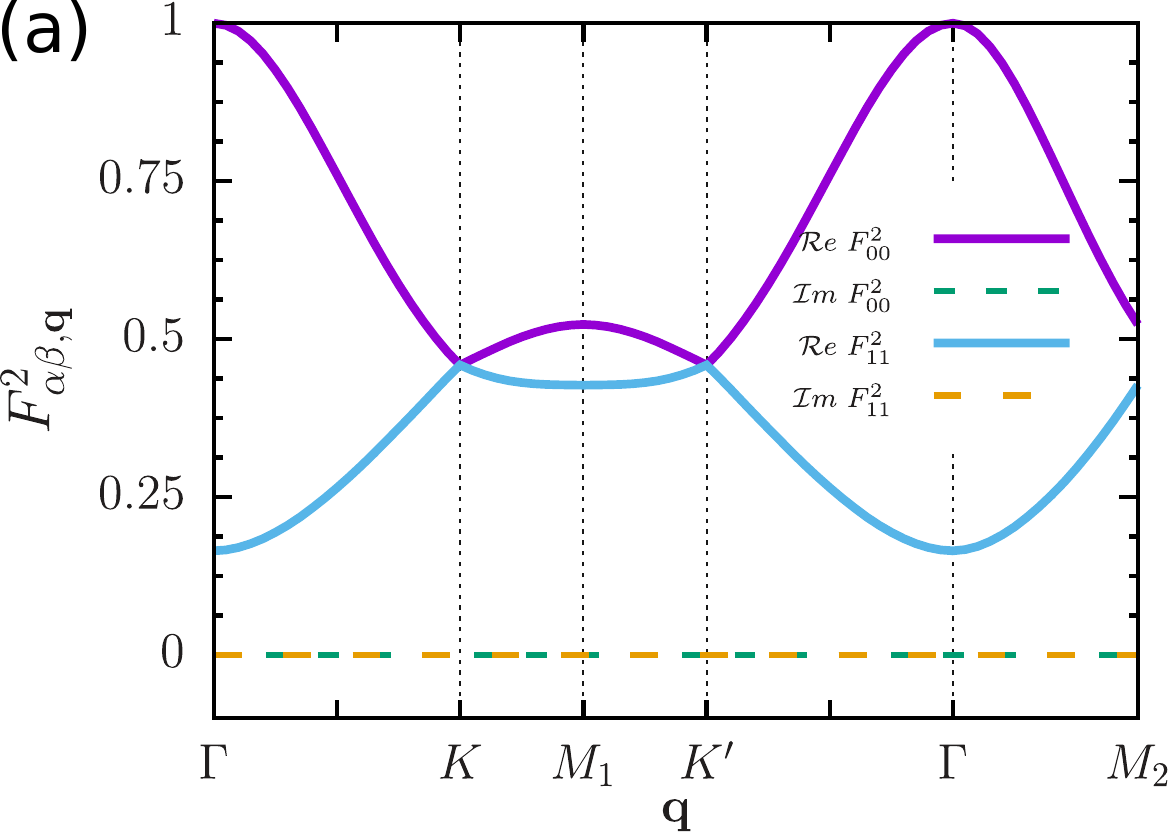} 
  \hskip0.4cm
  \includegraphics[width=5.5cm]{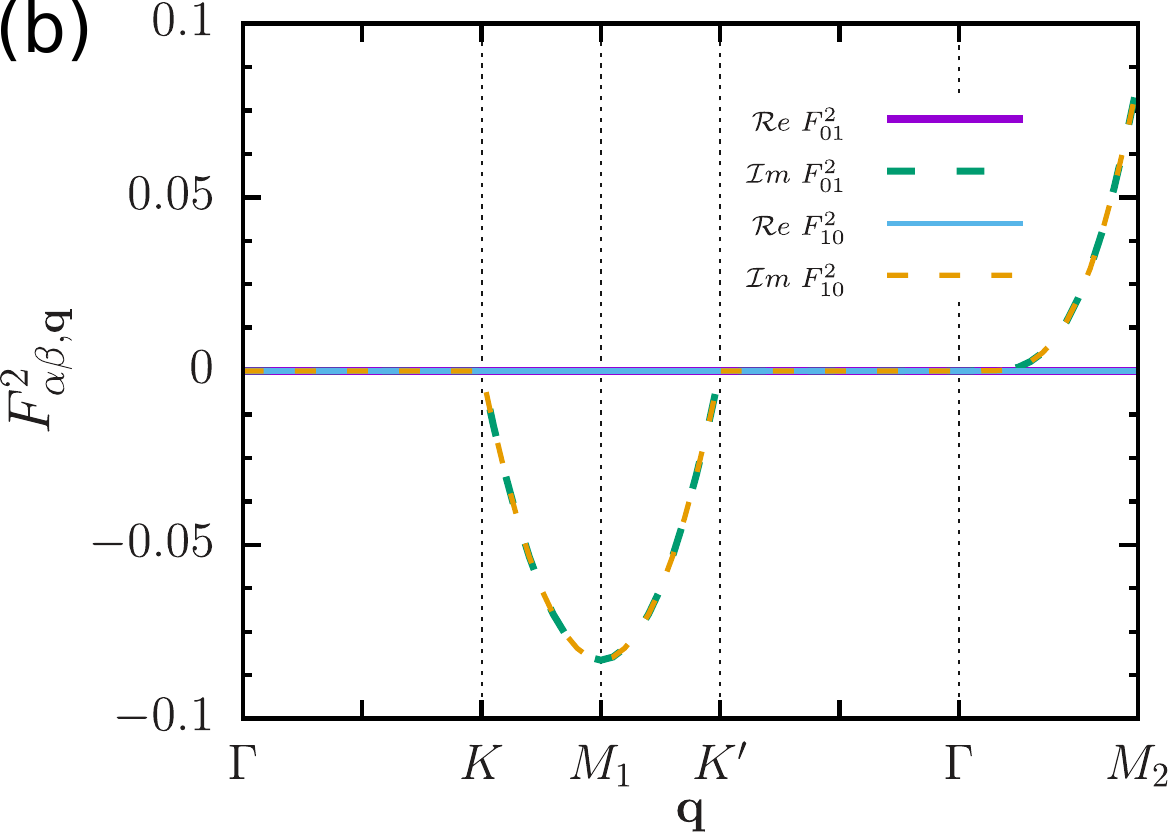}
  \hskip0.4cm
  \includegraphics[width=5.5cm]{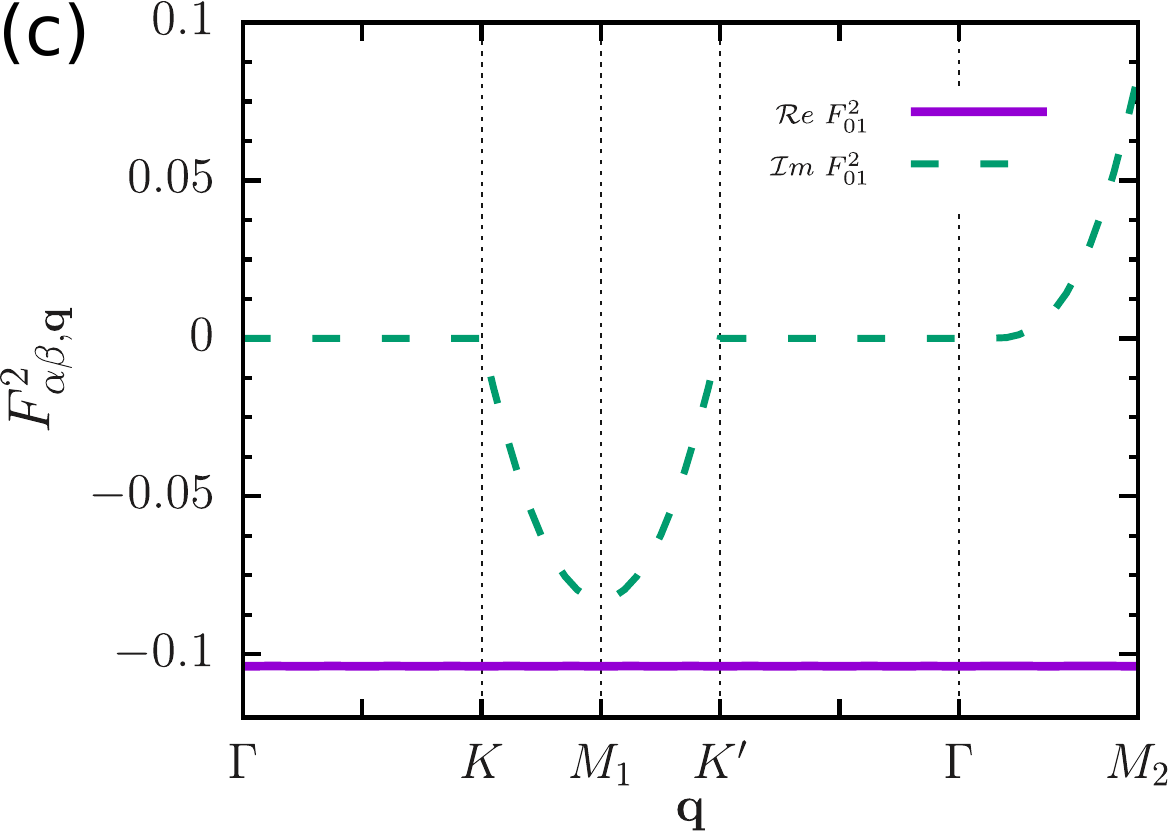}
}
\caption{The real (solid line) and imaginary (dashed line) parts 
 of the $F_{\alpha \beta,\bq}^2$  function [Eq.~\eqref{eq:F2}] for the
 Haldane model \eqref{eqHH2} in the 
 nearly-flat band limit \eqref{optimal-par}: 
 (a) $F_{00,\bq}^2$ and $F_{11,\bq}^2$ 
 (b) $F_{01,\bq}^2$ and $F_{10,\bq}^2$ for staggered on-site energy $M = 0$;
 (c) $F_{01,\bq}^2$ for staggered on-site energy $M = 0.2\, t_1$.}
\label{fig:F2}
\end{figure*}

\begin{widetext}
\section{Expressions of the $F_{\alpha \beta,\mathbf{q}}$ and 
            $\mathcal{G}_{\alpha \beta a \sigma}(\bk,\bq)$ functions}
\label{ap:functions}

In this section, we quote the expansions of the $F_{\alpha \beta,\mathbf{q}}$ and 
$\mathcal{G}_{\alpha \beta a \sigma}(\bk,\bq)$ functions in terms of
the coefficients \eqref{eqBs} that were derived in Ref.~\cite{doretto2015flat}.

The $F_{\alpha \beta,\mathbf{q}}$ function \eqref{eq:F2} is defined
in terms of the $g_\alpha(\bp,\bq)$ function \eqref{eq:ga}, which
is written in terms of the coefficients $u_\bk$ and $v_\bk$ of the
canonical transformation \eqref{eq:BogoTransf}. 
With the aid of Eq.~\eqref{eq:Bogocoef}, one shows that 
\begin{align}
 F^2_{\alpha \beta, \bq }  &=  \frac{1}{4} \sum_\bp  
       \left[ 1 + (-1)^{\alpha + \beta}\right] \left( 1 + \hat{B}_{3,\bp} \hat{B}_{3, \bp-\bq} \right) 
     -\left[ 1 - (-1)^{\alpha + \beta}\right]  \left( \hat{B}_{3,\bp} + \hat{B}_{3, \bp - \bp} \right) 
\nonumber \\
    &+ \left[ (-1)^\alpha + (-1)^\beta\right] \left( \hat{B}_{1,\bp} \hat{B}_{1, \bp-\bq}   + \hat{B}_{2,\bp} \hat{B}_{2, \bp-\bq} \right) 
     + i\left[ (-1)^\alpha - (-1)^\beta\right] \left( \hat{B}_{1,\bp} \hat{B}_{2, \bp-\bq}   - \hat{B}_{2,\bp} \hat{B}_{1, \bp-\bq} \right), 
\label{eq:ApF201}
\end{align}
with $\alpha,\beta = 0,1$ and $\hat{B}_{i,\bk} = B_{i,\bk} /|\mathbf{B}_\bk|$.

The $\mathcal{G}_{\alpha \beta a \sigma}(\bk,\bq)$ function, which determines 
the bosonic expression \eqref{eq:rhoBoson} of 
the projected density operator $\bar{\rho}_{a \sigma}(\bk)$, is defined as 
\begin{eqnarray}
\mathcal{G}_{\alpha \beta a \uparrow}( \mathbf{k} , \mathbf{q}) &=& - \sum_{ \mathbf{p}}
  \frac{G_a(\mathbf{p}, \mathbf{k})}{F_{\alpha\alpha, \mathbf{q}} F_{\beta\beta, \mathbf{k+q}}} g_{\alpha}(\mathbf{p-k}, \mathbf{q})
   g_{\beta}^*(\mathbf{p},\mathbf{k+q}),
\nonumber \\
\label{Gcal} \\
  \mathcal{G}_{\alpha \beta a \downarrow}( \mathbf{k} , \mathbf{q}) &=& +\sum_{ \mathbf{p}} 
  \frac{G_a(\mathbf{p-q}, \mathbf{k})}{F_{\alpha\alpha, \mathbf{q}} F_{\beta\beta, \mathbf{k+q}}} g_{\alpha}(\mathbf{p}, \mathbf{q}), 
    g_{\beta}^*(\mathbf{p},\mathbf{k+q}),
\nonumber
\end{eqnarray} 
where
\begin{equation}
    G_a(\mathbf{p}, \mathbf{q}) = \delta_{a,A}v_{\mathbf{p}-\mathbf{q}}^* v_{\mathbf{p}} 
                                                 + \delta_{a,B} u_{\mathbf{p}-\mathbf{q}}^* u_{\mathbf{p}}.
\label{eq:Ga}
\end{equation}
with $u_\bk$ and $v_\bk$ being the coefficients of the
canonical transformation \eqref{eq:BogoTransf}. 
It is then possible to show that
\begin{align}
\mathcal{G}_{\alpha \beta a \uparrow }( \mathbf{k} , \mathbf{q}) = 
  &-\frac{1}{8} [\delta_{a,A} + \delta_{a,B}(-1)^{\alpha + \beta}] \frac{1}{F_{\alpha, \mathbf{q}} F_{\beta , \mathbf{k+q}}} 
   \nonumber \\
  & \times\sum_p 1 -3(-1)^{a} \hat{B}_{3, \mathbf{p}} + \hat{B}_{3,\mathbf{p-q}} \hat{B}_{3, \mathbf{p+k}}
     + \hat{B}_{3,\mathbf{p-q}}\hat{B}_{3, \mathbf{p}} + \hat{B}_{3, \mathbf{p+k}}
        \hat{B}_{3, \mathbf{p}} -(-1)^{a} \hat{B}_{3, \mathbf{p-q}}\hat{B}_{3, \mathbf{p}} \hat{B}_{3, \mathbf{p+k}} 
   \nonumber \\
  &+ (-1)^{\alpha} [ \hat{B}_{1, \mathbf{p-q}} \hat{B}_{1, \mathbf{p}} + \hat{B}_{2, \mathbf{p-q}} \hat{B}_{2, \mathbf{p}} 
    + i(-1)^a (\hat{B}_{1, \mathbf{p-q}} \hat{B}_{2, \mathbf{p}} 
    - \hat{B}_{2,\mathbf{p-q}} \hat{B}_{1, \mathbf{p}})] [1-(-1)^a \hat{B}_{3,\mathbf{p+k}}] 
   \nonumber \\
  &+(-1)^{\beta} [\hat{B}_{1, \mathbf{p-q}} \hat{B}_{1, \mathbf{p+k}}
    + \hat{B}_{2, \mathbf{p-q}}\hat{B}_{2, \mathbf{p+k}} -i(-1)^{a}(\hat{B}_{1, \mathbf{p-q}} \hat{B}_{2, \mathbf{p+k}} 
     - \hat{B}_{2,\mathbf{p-q}}\hat{B}_{1, \mathbf{p+k}}][1 - (-1)^a \hat{B}_{3,\mathbf{p}}] 
  \nonumber \\
  &+(-1)^{\alpha + \beta}[\hat{B}_{1, \mathbf{p+k}} \hat{B}_{1,\mathbf{p}} + \hat{B}_{2, \mathbf{p+k}}\hat{B}_{2, \mathbf{p}}
    +i(-1)^a (\hat{B}_{1, \mathbf{p+k}} \hat{B}_{2, \mathbf{p}}
    -\hat{B}_{2, \mathbf{p+k}} \hat{B}_{1,
    \mathbf{p}})][1+(-1)^a\hat{B}_{3, \mathbf{p-q}}]. 
\label{Gcal2}
\end{align}
The expansion of the
$\mathcal{G}_{\alpha \beta a \downarrow}(\bk,\bq)$ function easily
follows from Eq.~\eqref{Gcal2}, since 
$\mathcal{G}_{\alpha \beta a \downarrow}(\bk,\bq) =
-\mathcal{G}^*_{\alpha \beta a \uparrow}(-\bk,-\bq)$.

\end{widetext}

\section{Details about the bosonization scheme}
\label{ap:BosoDetails}

In this section, we provide some details about the definition of the
boson operators \eqref{eq:bosons} and discuss the differences between
the application of the bosonization formalism for the Haldane
model and the square lattice $\pi$-flux model \cite{doretto2015flat}. 
Indeed, the application of the bosonization scheme
\cite{doretto2015flat} for the Haldane model on a honeycomb lattice
requires further approximations as compare to the case of the
square lattice $\pi$-flux model.

\begin{figure*}[t]
\centerline{
  \includegraphics[width=5.5cm]{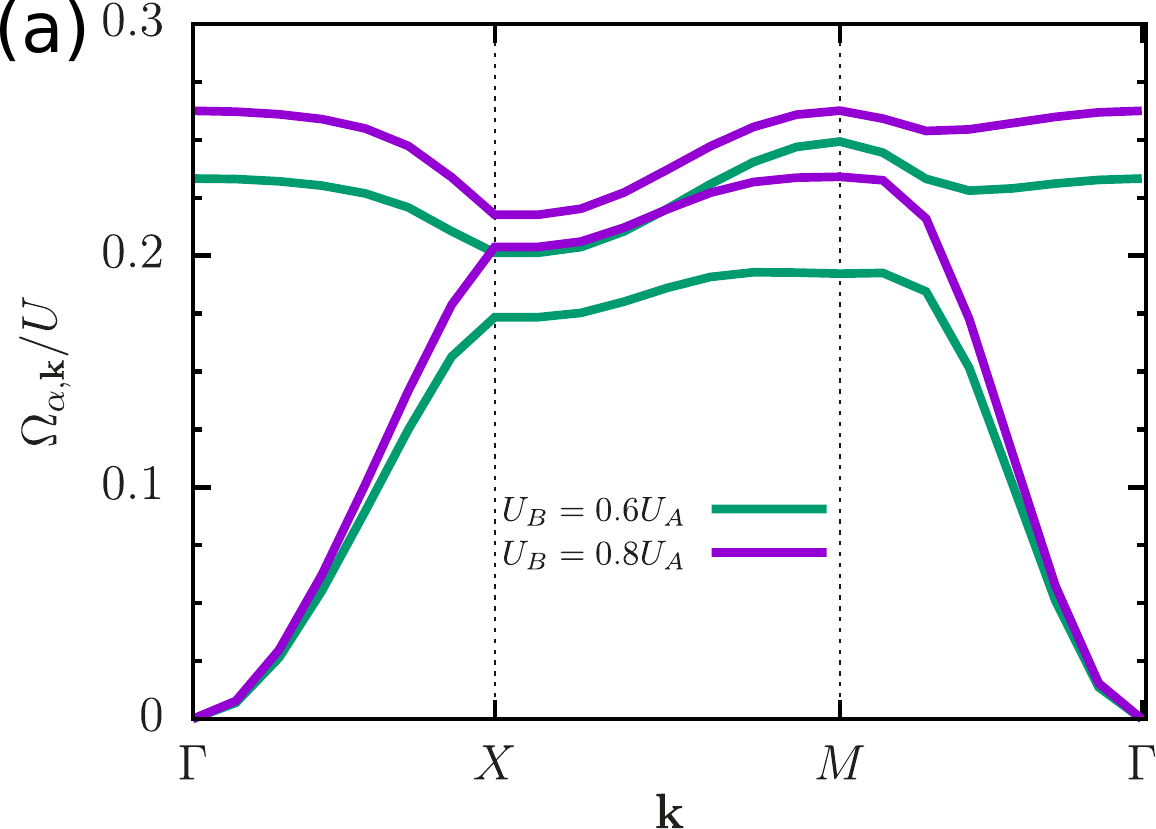}
  \hskip0.5cm
  \includegraphics[width=5.5cm]{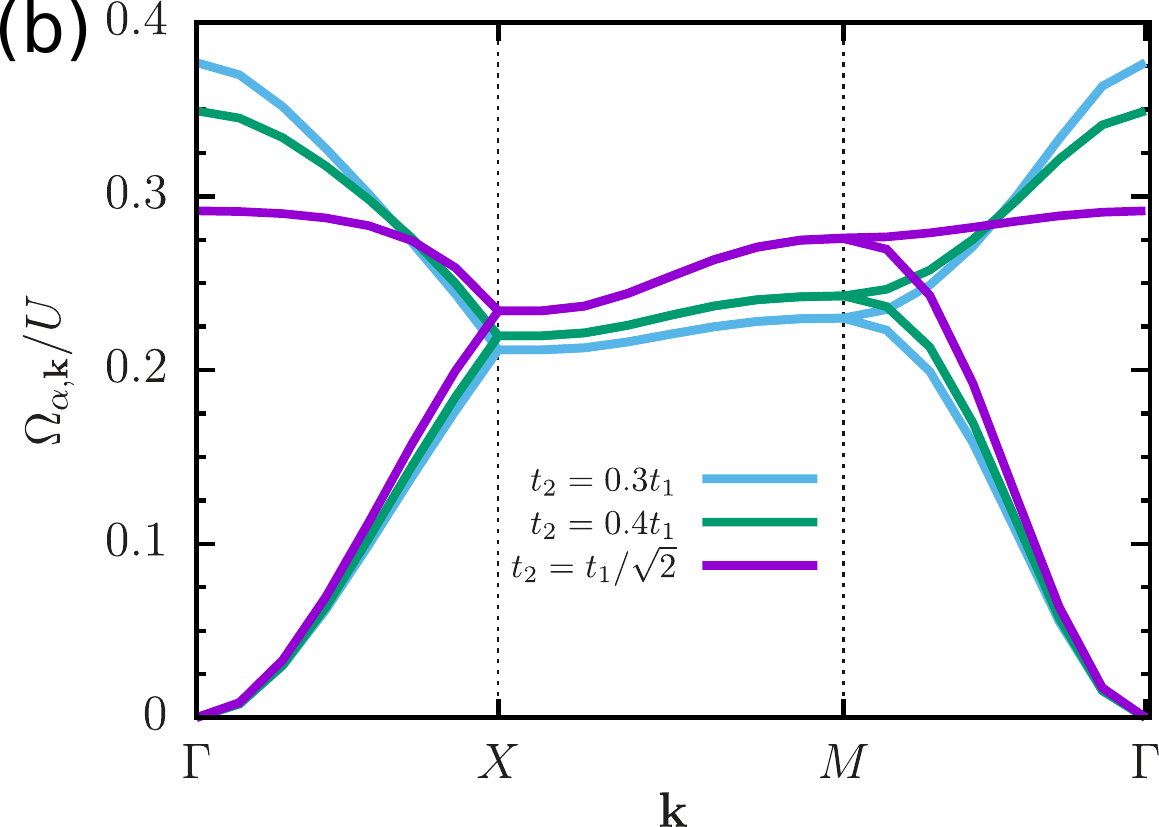}
  \hskip0.5cm
  \includegraphics[width=5.5cm]{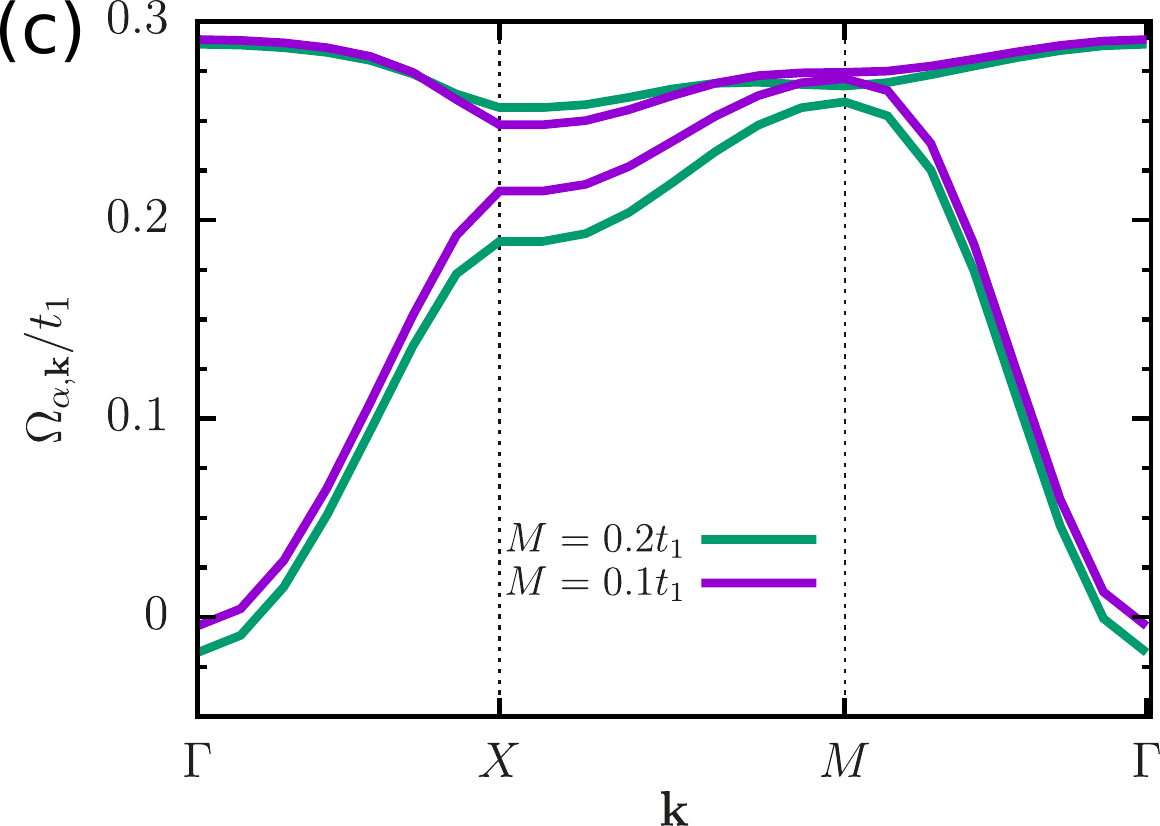}
}
\caption{Spin-wave excitation spectra \eqref{omega-b} for the
  flat-band ferromagnetic phase of the square lattice $\pi$-flux model
  (see Ref.~\cite{doretto2015flat} for details):  
  (a) Next-nearest-neighbor hopping amplitudes $t_2 = 1/\sqrt{2}$ and
  on-site repulsion energy $U_B = 0.8U_A = 0.8\, U$ (magenta) and
  $U_B = 0.6U_A = 0.6\, U$ (green);
  (b) $U_A = U_B = U$ and
  $t_2 = 1/\sqrt{2}$ (magenta),
  $0.4$ (green), and
  $0.3\, t_1$ (blue);
  (c) $t_2 = 1/\sqrt{2}$, $U_A = U_B = U$, and
  staggered on-site energy 
  $M = 0.1$ (magenta) and
  $0.2\, t_1$ (green). 
}
\label{fig:MassSquare}
\end{figure*}

As mentioned in Sec.~\ref{sec:boso}, the boson operators \eqref{eq:bosons}
are defined in terms of projected spin operators in momentum
space:
\[
   b_{\alpha,\mathbf{q}}  \propto \bar{S}_{-\mathbf{q},\alpha}^{+} 
   \quad\quad {\rm and} \quad\quad
   b_{\alpha,\mathbf{q}}^{\dagger} \propto \bar{S}_{\mathbf{q},\alpha}^{-}.
\]
In terms of the fermion operators $c^\dagger_{\bk\,\sigma}$ and $c_{\bk\,\sigma}$ 
(associated with the lower noninteracting band $c$),
the commutator between the projected spin operators 
$\bar{S}_{-\mathbf{q},\alpha}^{+}$ and
$\bar{S}_{\mathbf{q},\alpha}^{-}$ reads 
\begin{align}
[\bar{S}^+_{\bq,\alpha}, \bar{S}^-_{\bq',\beta}] =& \sum_\bp  
            \left[ g_\alpha(\bp-\bq',\bq)g_\beta(\bp,\bq')
                c^\dagger_{\bp-\bq-\bq'\,\uparrow}c_{\bp\,\uparrow}
           \right.
\nonumber \\
 &-  \left.    g_\alpha(\bp,\bq)g_\beta(\bp-\bq,\bq')
             c^\dagger_{\bp-\bq-\bq'\,\downarrow}c_{\bp\,\downarrow}   \right],
\label{comut-ss} 
\end{align}
with the $g_\alpha(\bp,\bq)$ function giving by Eq.~\eqref{eq:ga}.
One sees that Eq.~\eqref{comut-ss} is different from the canonical
commutation relation \eqref{eq:BComutations} for boson operators.
However, as long as we are close to the ferromagnetic state \eqref{eq:FM},
i.e., the number of particle-hole pair excitations is small, one can
assume that 
\begin{align}
 c^\dagger_{\bp-\bq\,\uparrow}c_{\bp\,\uparrow} &\approx 
  \langle {\rm FM} | c^\dagger_{\bp-\bq\,\uparrow}c_{\bp\,\uparrow} | {\rm FM} \rangle
       =  \delta_{\bq,0},
\nonumber \\
   c^\dagger_{\bp-\bq\,\downarrow}c_{\bp\,\downarrow} &\approx
  \langle {\rm FM} | c^\dagger_{\bp-\bq\,\downarrow}c_{\bp\,\downarrow} | {\rm FM} \rangle
  =  0,
\label{assumption}
\end{align}
and therefore, the commutator \eqref{comut-ss} now reads
\begin{equation}
	  [\bar{\mathbf{S}}_{\mathbf{q},\alpha}^{+} ,
          \bar{\mathbf{S}}_{\mathbf{q}',\beta}^{-} ] \approx  \delta_{\mathbf{q},\mathbf{-q}'} F_{\alpha \beta,\mathbf{q}}^2,
\label{comut-ss-aprox}
\end{equation}
where the $F_{\alpha \beta,\mathbf{q}}$ function is defined by Eq.~\eqref{eq:F2}. 
The commutator \eqref{comut-ss-aprox} indicates that the particle-hole
pair excitations \eqref{eq:Sexcitation} can be treated approximated as bosons. 
The bosonization formalism for flat-band Chern insulators
\cite{doretto2015flat} is indeed based on the assumption \eqref{assumption}.

For the square lattice $\pi$-flux model \cite{doretto2015flat}, it was found that the condition 
\eqref{conditionF} holds, and therefore,  Eq.~\eqref{comut-ss-aprox}
allows us to define two sets of {\sl independent} boson operators
$b_0$ and $b_1$ as done in Eq.~\eqref{eq:bosons}. On the other hand,
for the Haldane model \eqref{eqHH2} on the honeycomb lattice, the condition \eqref{conditionF} 
is not fulfilled, as exemplified in Figs.~\ref{fig:F2}(a) and (b) for
the nearly-flat band limit \eqref{optimal-par}. Since
$F_{\alpha \alpha,\mathbf{q}}^2$ are real quantities, 
the imaginary parts of $F_{01,\mathbf{q}}^2$ and $F_{10,\mathbf{q}}^2$
are finite only in the vicinity of the $M_1$ and $M_2$ points, 
and $|F_{01,\mathbf{q}}^2|$, $|F_{10,\mathbf{q}}^2| < |F_{\alpha \alpha,\mathbf{q}}^2|$,
we assume that, for the Haldane model,
bosons operators $b_0$ and $b_1$ can still be
defined by Eq.~\eqref{eq:bosons}  and that they constitute two sets of
independent boson operators.

A second important distinction between the Haldane and square lattice
$\pi$-flux models is associated with the determination of the bosonic
representation of operators written in terms of the fermion operators $c_{\bq\,\sigma}$,
such as the projected electron density operator $\bar{\rho}_{a\sigma}(\bk)$ 
[Eq.~\eqref{eq:rhoBoson}]. As discussed in Sec.~III.B from Ref.~\cite{doretto2015flat}, such a
procedure is based on the fact that one can {\sl define} the product
of fermion operators $c^\dagger_{\bp-\bq\,\downarrow}c_{\bp\,\uparrow}$
in terms of the boson operators $b_\alpha$, i.e.,
\begin{equation}
  c^\dagger_{\bp-\bq\,\downarrow}c_{\bp\,\uparrow}  \equiv \sum_\beta  
                 h_\beta(\bp,\bq) b^\dagger_{\beta,\bq}.
\label{cc1}
\end{equation}
For the square lattice $\pi$-flux model, where the condition
\eqref{conditionF} holds, it is easy to see that the $h_\beta(\bp,\bq)$ 
function is given by
\begin{equation}
  h_\beta(\bp,\bq) = \frac{1}{F_{\beta\beta,\bq}} g^*_\beta(\bp,\bq),
\label{h-function}
\end{equation}
since the substitution of Eqs.~\eqref{cc1} and \eqref{h-function} into
\eqref{eq:bosons} yields
\begin{align}
   b_{\alpha,\mathbf{q}}^{\dagger} &= \frac{1}{F_{\alpha\alpha,\mathbf{q}}}  
                \sum_\beta \left[ \sum_{\mathbf{p}} h_\beta(\bp,\bq) g_{\alpha} (\mathbf{p}, \mathbf{q})  \right]   
               b_{\beta,\mathbf{q}}^{\dagger}  
\nonumber \\
        &= \frac{1}{F_{\alpha\alpha,\mathbf{q}}}  
                \sum_\beta    \frac{ \delta_{\alpha, \beta} F^2_{\alpha\beta,\bq}}{F_{\beta\beta,\bq}}
               b_{\beta,\mathbf{q}}^{\dagger} = b_{\alpha,\mathbf{q}}^{\dagger}, 
\label{identity-b}
\end{align}
see also Eq.~\eqref{eq:F2}. 
For the Haldane model on the honeycomb lattice, the choice
\eqref{h-function} for the $h_\beta(\bp,\bq)$ function seems to be not
appropriated, since the condition \eqref{conditionF} is no longer
valid. Due to the involved expansion of the $g_\alpha(\bp,\bq)$
function in terms of the coefficients \eqref{eqBs} (not shown here),
it is difficult to determine an  $h_\beta(\bp,\bq)$
function such that the identity \eqref{identity-b} is satisfied.
Therefore, based on the same assumptions considered in the definition
of the boson operators $b_\alpha$ and discussed in the previous
paragraph, we also assume that Eq.~\eqref{h-function} [and
consequently Eq.~\eqref{eq:rhoBoson}] holds for the
Haldane model.

As discussed in Sec.~\ref{sec:spin-wave}, one important 
consequence of the fact that the condition \eqref{conditionF} is not
fulfilled for the Haldane model is that the coefficients 
$\bar{\omega}^{01}_\bq$ and  $\bar{\omega}^{10}_\bq$ 
[Eq.~\eqref{eq:omegaBar}] are finite and the coefficients \eqref{eq:Epsilon}
obey the relation $\epsilon^{01}_\bq = -\epsilon^{10}_\bq$, 
yielding a non-Hermitian quadratic boson Hamiltonian \eqref{H42}.
It is indeed easy to understand the relation between these results
once we compare the integrand of Eq.~\eqref{eq:F2} with the ones of
Eqs.~\eqref{eq:omegaBar} and \eqref{Gcal}: Notice that all of
them depend on the product $g_\alpha(x,y)g^*_\beta(x',y')$; 
for $\alpha \not= \beta$, additional terms might be included in the 
$\mathcal{G}_{\alpha \beta a \sigma}(\bk,\bq)$ function, yielding  
$\epsilon^{01}_\bq = -\epsilon^{10}_\bq$.
In principle, the condition $\epsilon^{01}_\bq = (\epsilon^{10}_\bq)^*$ 
could be restore, once an appropriated choice for the
$h_\beta(\bp,\bq)$ function were done such that the 
identity \eqref{identity-b} is now verified.

\section{Square lattice $\pi$-flux model}
\label{ap:square}

In this section, we present additional results derived within the
bosonization formalism for the flat-band ferromagnetic phase of the
topological Hubbard model on a square lattice, 
whose noninteracting limit is given by the $\pi$-flux model,
previously studied in  Ref.~\cite{doretto2015flat}. 
We follow the lines of  Secs.~\ref{sec:spin-wave} and
\ref{sec:dispersiviness} and find that the spin-wave spectra of both 
square lattice $\pi$-flux and Haldane models display the same
features.

\begin{figure*}[t]
\centerline{
  \includegraphics[width=7.0cm]{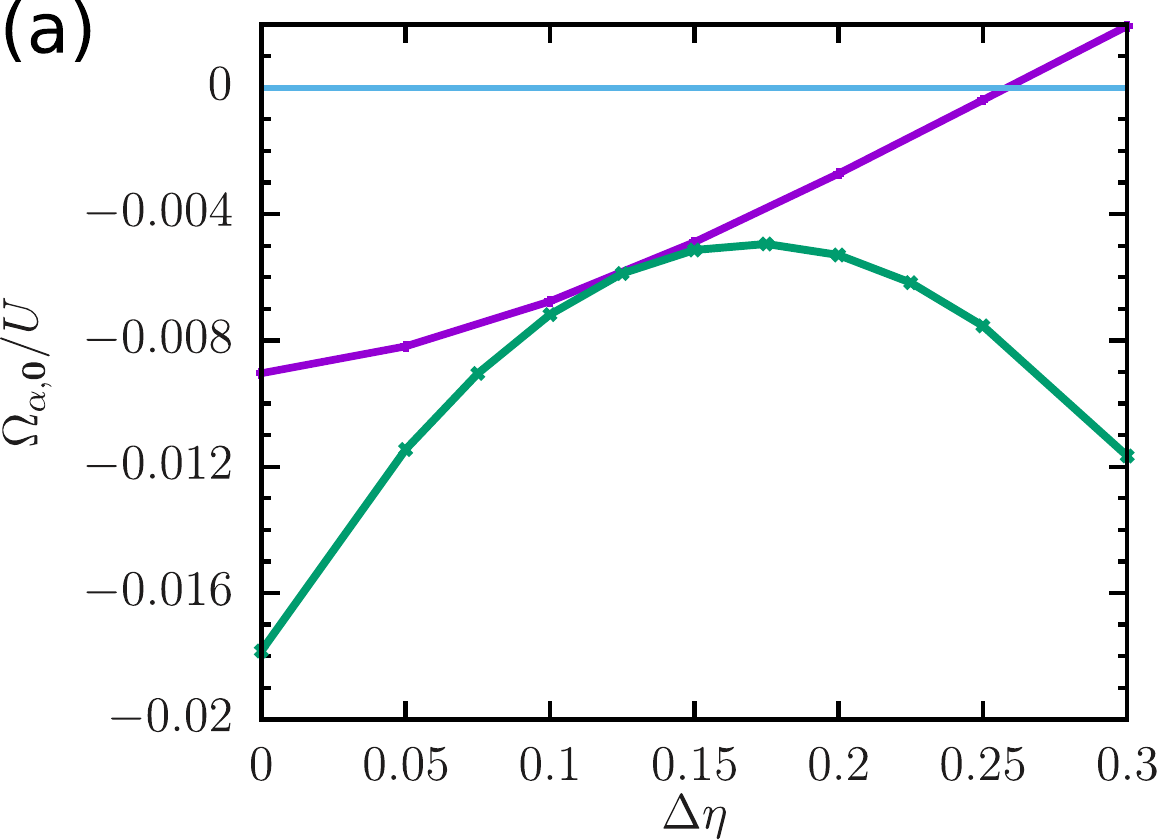}
  \hskip1.2cm
  \includegraphics[width=7.0cm]{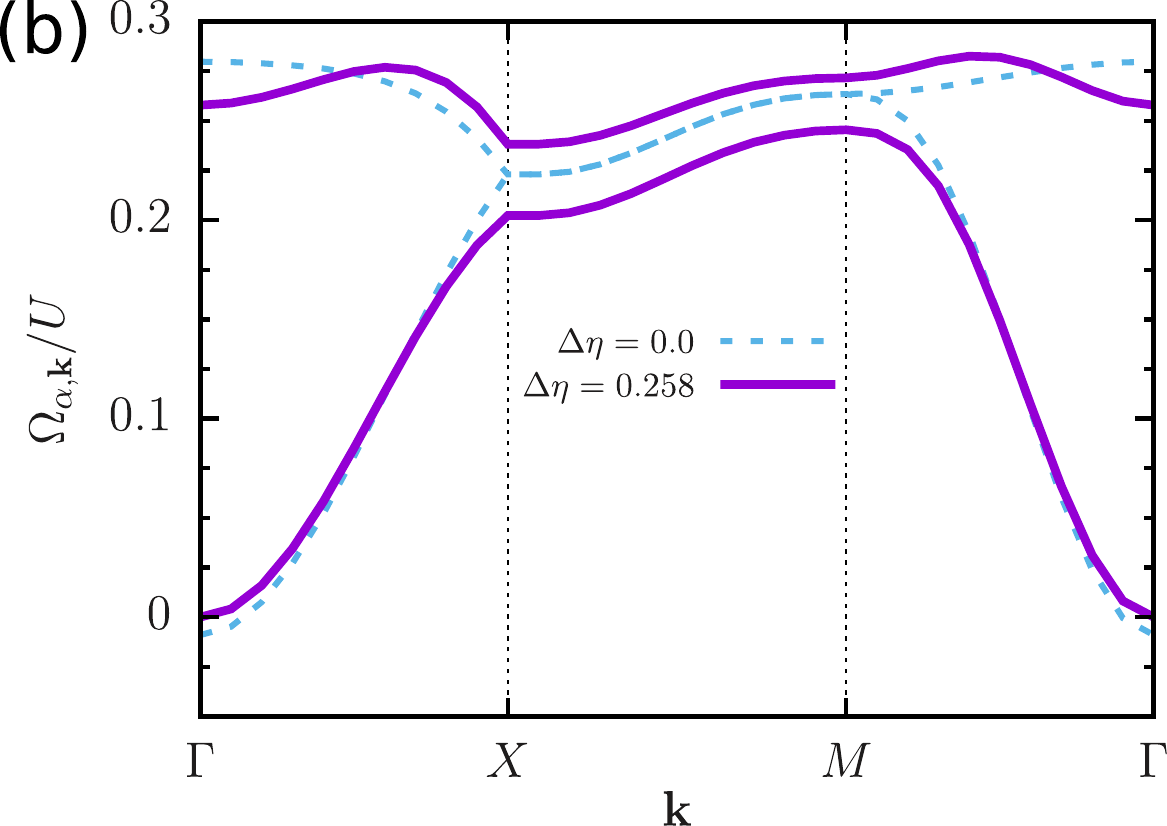} 
}
\caption{(a) The energy of the Goldstone mode for the square lattice
  $\pi$-flux model with one-site staggered energy $M=0.2\, t_1$ 
  in terms of $\Delta\eta = \eta - \pi/4$
  determined with the dispersion relations 
  \eqref{omega-b} (solid green line) and
  \eqref{omega-b2} (solid magenta line).
  (b) Dispersion relation \eqref{omega-b2} for 
  $\Delta\eta = 0$ (dashed blue line) and
  $\Delta\eta = 0.258$ (solid magenta line).}

\label{fig:GammaXangle}
\end{figure*}

Figure~\ref{fig:MassSquare}(a) shows the spin-wave spectrum \eqref{omega-b} 
for the nearly flat-band limit of the square lattice $\pi$-flux model
(which corresponds to the configuration with the next-nearest-neighbor
hopping amplitude $t_2 = t_1/\sqrt{2}$) and on-site repulsion energies 
$U_B = 0.8\, U_A = 0.8\, U$ and $U_B = 0.6\, U_A = 0.6\, U$.
A comparison with the spin-wave spectrum obtained for the homogeneous
case $U_B = U_A = U$ (see Fig.~4 from Ref.~\cite{doretto2015flat} and
Fig.~\ref{fig:MassSquare}(b))
indicates that the energies of the excitations decrease with $\Delta U$
and an energy gap opens at the border of the first Brillouin zone
(the $X$-$M$ line). Such features where also found for the
Haldane-Hubbard model, see Fig.~\ref{figEspectro1}.
Importantly, for the square lattice $\pi$-flux model, the decay rates of
the spin-wave excitations vanish.

The effects of a decreasing of the flatness ratio of the
noninteracting bands due to the variation of the next-nearest-neighbor
hopping amplitude $t_2$ (see Fig.~3 from Ref.~\cite{doretto2015flat}
for details) are shown in Fig.~\ref{fig:MassSquare}(b). Apart from a
renormalization of the excitation energies, the spin-wave spectrum
\eqref{omega-b} display the same features of the nearly flat-band
limit, similar to the behaviour found for the Haldane-Hubbard model, 
see Fig.~\ref{fig:Omega-phi}.

Finally, the effects of a finite staggered on-site energy $M$ are
presented in Fig.~\ref{fig:MassSquare}(c). 
Since the kinetic contribution $\bar{\omega}^{\alpha,\alpha}_\bq$ is
quite small it is not considered.
In addition to open an energy gap at the first Brillouin
zone border, a finite $M$ also decreases the lower branch energies in
the vicinity of the $\Gamma$ point, which indicates an instability of
the flat-band ferromagnetic phase, see also Fig.~\ref{fig:Omega-Mass}(a).

As discussed in Sec.~\ref{sec:summary}, the instability of the
flat-band ferromagnetic phase in the presence of a finite staggered
on-site energy $M$ might be an artefact of the bosonization formalism
related to the kinetic contribution \eqref{eq:omegaBar}.
Although it is not clear yet how to properly include in the effective
boson model \eqref{heffective} the explicitly effects of the dispersion of the
noninteracting bands, we check weather a modification of the 
boson operators \eqref{eq:bosons} definition could restore the
Goldstone mode. 
In the following, we briefly summarize such a possible procedure and
apply it for the square lattice $\pi$-flux model.

The definition of the boson operators \eqref{eq:bosons} is based on
the linear combination \eqref{eq:SprojAlpha} of the projected spin
operators $\bar{S}_{\mathbf{q}, A/B}$. Since a finite $M$ introduces
an offset in the energies of the sites associated with the
sublattices $A$ and $B$, instead of Eq.~\eqref{eq:SprojAlpha}, 
one should consider the generalized form
\begin{align}
 \bar{S}_{\mathbf{q}, 0}^{\pm} &= \frac{\sqrt{2}}{2} \left( \cos(\eta)\bar{S}_{\mathbf{q}, A}^{\pm} 
                                       + \sin(\eta) \bar{S}_{\mathbf{q}, B}^{\pm} \right), 
\nonumber \\
\label{eq:NewS} \\
 \bar{S}_{\mathbf{q}, 1}^{\pm} &= \frac{\sqrt{2}}{2} \left( \sin(\eta)\bar{S}_{\mathbf{q}, A}^{\pm} 
                                       - \cos(\eta) \bar{S}_{\mathbf{q}, B}^{\pm} \right), 
\nonumber
\end{align}
where the linear combination \eqref{eq:SprojAlpha} can be obtained by
choosing the parameter $\eta = \pi/4$. 
In this case, the expansion of the $F^2_{\alpha\beta,\bq}$ function
\eqref{eq:F2} in terms of the coefficients \eqref{eqBs} now reads
\begin{widetext}
\begin{align}
 F_{\alpha \alpha , \mathbf{q}}^2 &= \frac{1}{4} \sum_{\mathbf{k}}
 [1+\hat{B}_{3 \mathbf{k}} \hat{B}_{3 \mathbf{k-q}}  ]  -
 (-1)^{\alpha}\frac{\cos(2 \eta) }{4} [ \hat{B}_{3
   \mathbf{k}}+\hat{B}_{3 \mathbf{k-q}}] + (-1)^{\alpha}\frac{\sin(2
   \eta) }{4} [\hat{B}_{1 \mathbf{k}} \hat{B}_{1 \mathbf{k-q}} +
 \hat{B}_{2 \mathbf{k}} \hat{B}_{2 \mathbf{k-q}}],  
\nonumber \\
\label{newF2}\\
 F_{\alpha\beta , \mathbf{q}}^2 &= \sum_{\mathbf{k}} -\frac{\sin(2 \eta) }{4}
  [\hat{B}_{3 \mathbf{k}} + \hat{B}_{3 \mathbf{k-q}}] - \frac{\cos(2
  \eta) }{4} [\hat{B}_{1 \mathbf{k}} \hat{B}_{1 \mathbf{k-q}} +
  \hat{B}_{2 \mathbf{k}} \hat{B}_{2 \mathbf{k-q}}]  - (-1)^\alpha
  \frac{i}{4}[\hat{B}_{2 \mathbf{k}} \hat{B}_{1 \mathbf{k-q}} -
  \hat{B}_{1 \mathbf{k}} \hat{B}_{2 \mathbf{k-q}}],
\nonumber
\end{align}
\end{widetext}
with $\alpha \not= \beta$, and the $g_\alpha(\bp,\bq)$ function
\eqref{eq:ga} is now given by
\begin{align}
  g_{\alpha}(\mathbf{p}, \mathbf{q}) &= G_A(\mathbf{p}, \mathbf{q})
  [\cos(\eta) (1-\alpha) + \alpha \sin(\eta)] 
\nonumber \\  
  &+  G_B(\mathbf{p},
  \mathbf{q}) [\sin(\eta) (1-\alpha) - \alpha \cos(\eta)],
\label{newga}
\end{align}
with the $G_a(\mathbf{p}, \mathbf{q})$ function defined by Eq.~\eqref{eq:Ga}.
Eqs.~\eqref{newF2} and \eqref{newga} implies that the expansion \eqref{Gcal2} of
the $\mathcal{G}_{\alpha \beta a \sigma}(\bk,\bq)$ function in terms
of the coefficients \eqref{eqBs} is also modified. 
Importantly, the modified $F^2_{\alpha\beta,\bq}$ function \eqref{newF2}  
with $\eta \not= \pi/4$ implies that the condition \eqref{conditionF}
is no longer valid for the square lattice $\pi$-flux model.

Figure~\ref{fig:GammaXangle}(a) shows the energy of the Goldstone
mode in terms of $\Delta \eta = \eta - \pi/4$ determined with both
dispersion relations \eqref{omega-b} and \eqref{omega-b2}
for on-site staggered energy $M = 0.2\, t_1$. 
One notices that it is possible to find a parameter $\eta$ such that 
the Goldstone mode is restored only if the coefficients
$\epsilon^{01}_\bq$ and $\epsilon^{10}_\bq$ [Eq.~\eqref{eq:Epsilon}]
are neglected. The spin-wave spectrum \eqref{omega-b2} for optimal
$\Delta \eta = 0.258$ is display in Fig.~\ref{fig:GammaXangle}(b).


\end{document}